# Writing Into Water

*Nadir Möller[†], Lukas Hecht[†], Ran Niu[‡], Benno Liebchen[\*], Thomas Palberg[\*]*

[†]These authors contributed equally to this work.

[‡]Present address: Huazhong University of Science and Technology, Luoyu Road 1037, Wuhan 430074, China.

N. Möller, R. Niu, T. Palberg

Institut für Physik

Johannes Gutenberg Universität

Staudinger Weg 7

55128 Mainz, Germany

E-mail: palberg@uni-mainz.de

L. Hecht, B. Liebchen

Institute for Condensed Matter Physics

Department of Physics

Technische Universität Darmstadt

Hochschulstr. 8

64289 Darmstadt, Germany

E-mail: benno.liebchen@pkm.tu-darmstadt.de

**Abstract**

Writing is an ancient communication technique dating back at least 30,000 years. While even sophisticated contemporary writing techniques hinge on solid surfaces for engraving or the deposition of ink, writing within a liquid medium requires a fundamentally different approach. We here demonstrate writing of lines, letters, and complex patterns in water by assembling lines of colloidal particles. Unlike established techniques for underwater writing on solid substrates, these lines are fully reconfigurable and do not require any fixation onto the substrate. Exploiting gravity, we roll an ion-exchange bead (pen) across a layer of sedimented colloidal particles (ink). The pen evokes a hydrodynamic flow collecting ink-particles into a durable, high-contrast line along its trajectory. Deliberate substrate-tilting sequences facilitate pen-steering and thus to draw and write. We complement our experiments with a minimal model that quantitatively predicts the observed parameter dependence for writing in fluids and highlights the generic character of writing by line-assembly. Overall, our approach opens a versatile route for writing, drawing, and patterning fluids – even at the micro-scale.

# 1. Introduction

Traditional writing techniques comprise carving and engraving as well as printing and writing with ink. Earliest human drawings date back some 30,000 years, possibly even much longer.[1] As visible language, writing appeared in the Middle East between 3,400 and 2,600 before the Common Era.[2,3] These techniques continue to coexist as means of storing and transporting information, nowadays accompanied by various multimedia techniques for displaying.[4] In addition, various novel techniques extend and complement these traditional techniques, including (electron) lithography, optical tweezing, direct printing, or force microscopic manipulation.[5–12] Remarkably, the size of glyphs and letters covers the range from a few hundred meters[13] down to the atomic scale[14] and even below.[15] In the more classical approaches, one creates a local, line-shaped variation of the material density in or on an extended substrate acting as background: A line is carved out or some ink is deposited. A solid substrate stabilizes the density variation by strong intermolecular forces, keeping it in shape. The same principle has been applied to write on surfaces submerged in a fluid. For instance, scanning probe lithography was used to carve or deposit lines within or onto self-assembled monolayers submerged in fluids containing suitable chemicals.[16,17] In addition, sophisticated micron-sized structures have been printed using two-photon polymerization.[18,19] UV-polymerization and crosslinking was also used to write on a solid surface within a liquid starting from a dispersion of reactive chemicals to manufacture patterns with superb thermoresponsive mechanical properties.[20] There are now even commercial scuba diver slates available for under-water writing on a substrate. Importantly, however, all these approaches still rely on a substrate (i) for fixing the written structures and (ii) for providing mechanical support. In contrast, *writing into a fluid* requires a mechanism that does not depend on such localization measures. The mechanism must also be intrinsically robust against rapid line dispersion, which would cause short lifetimes of any drawn lines. In fact, even in a quiescent (convection-free) fluid, the moving pen would transfer kinetic energy to the fluid, provoking line dispersion by locally created eddies. While such local eddies are rather unimportant when the pen is much smaller than the written letters, as, e.g., in skywriting,[21] the creation of fine, durable, and freely-floating lines remains challenging. In fact, to write fully reconfigurable lines into a liquid at the microscale, an approach fundamentally different from underwater ink deposition or line carving and a new type of micro-pen are required. To develop such an approach, we exploit the following ideas.

Incidentally, a mobile fluid offers an alternative way of writing lines by particle transport towards a prescribed pattern. Imagine to start from a homogenous density of ink particles in a quiescent fluid and to use a pen which attracts the ink particles towards itself and/or its trajectory. If the resulting accretion process is sufficiently efficient and fast as compared to the subsequent dispersion of the ink particles, an increase of ink density may result past the pen and a line is written. As key ingredients, this approach requires a sufficient range of the particle-transporting attraction, a slow line dispersion, and a suitable way of pen steering.

To meet the first requirement, directed transport of colloidal particles by chemical, thermal, or light-intensity gradients can be exploited. A key example are phoretic effects, where, in general, the imposed gradient leads to a difference in chemical potential along a particle surface and drives a slip flow of the adjacent fluid along the surface, which, in turn, evokes directed motion of the particle.[22–26] Using a large chemical-loaded "beacon" falling under gravity, Banerjee et al. created a colloidal over-density evolving along the trajectory within a few minutes.[27] Here, colloidal motion relied on the local strength and direction of the gradient of electrolyte concentration. While these pioneering experiments demonstrate the possibility of writing freely buoyant lines within a fluid, they lack the option of deliberate pen steering.

In this work, we demonstrate a generic method for writing lines and letterings into a liquid rather than onto a solid. This method uses an ion-exchange resin bead (IEX) as fully steerable micro-pen and exploits the presence of a solid substrate only for line assembly but not for fixing the ink. Instead, lines are written near a substrate but are not attached to it yielding freely floating long-lived lines which can be reconfigured and allow us to recycle the ink for writing new lines before optionally fixing them to the substrate.[20,28,29] To achieve this, we exploit an effective way to transport colloids along extended surfaces: The ion-exchange resin bead (pen) evokes a so-called diffusio-osmotic (DO) flow.[22,30,31] Such a flow emerges because the ion-exchange process induces a nonuniform concentration profile which causes a stress (force) on the solvent within the interfacial layer of the substrate resulting in a solvent flow towards the IEX.[32] This flow then advects colloidal tracers towards the IEX, which can be viewed as an effective attraction between the IEX and the tracers.[33] In the past, DO flows have been successfully employed with fixed sources,[32–38] self-propelling sources,[39,40] and combinations of sources and sinks[41] to create centrally symmetric or asymmetric assemblies of tracer particles at the source. Here, we exploit this mechanism to dynamically assemble a line of tracer particles (ink) in the wake of a moving ion-exchange resin bead. Once assembled, those tracers will disperse only very slowly by thermal diffusion in the eddy-free fluid. Moreover, the ongoing pen-induced DO flows focus ink particles towards the center of the written lines supporting their durability and sharpness.

Our results exemplify a generic pathway for writing and drawing fine, free-floating but durable lines in a liquid medium. Our approach is modular and allows combinations of different inks, drives, steering, and, optionally, fixation techniques. This could be used in the future for structuring liquids with deliberate line-based patterns, decorate and thus visualize chemical tracers, or create desired initial states for future colloid experiments.

## 2. Results

### 2.1. Writing of straight lines

Our experimental approach is sketched in **Figure 1**a. Samples are placed on a programmable, motorized stage, allowing for tilting by $\vartheta$ and rotation around the optical axis (OA) by $\varphi$, and are observed at moderate magnification (10× or 20×) using an inverted microscope. A dilute suspension of weakly charged, micro-sized silica spheres (Si832, tracers, T), is left to settle under gravity in a slit cell of 500 μm height onto a charged glass substrate. To this dilute homogeneous monolayer, we add a larger bead of cationic ion-exchange resin (C-IEX45, IEX). It exchanges residual cations ($c \approx 10^{-8}$ mol L$^{-1}$) for protons and thus acts as mobile proton source.[32] The high proton mobility quickly establishes a large-scale pH-field.[42] Tilting the substrate in $x$-direction by an angle $\vartheta$ lets this IEX roll straight across the substrate with a velocity $v_{IEX} = (gV\Delta\rho/\gamma) \sin\vartheta$ (red arrow in Figure 1a and Figure S1a, Supporting Information). Here, $g$ is the gravitational acceleration, and $V$, $\Delta\rho$, and $\gamma$ are the volume, density mismatch, and the drag coefficient of the IEX. Tracers also roll, but at negligibly small velocities. The rolling IEX (pen) remains embedded in its self-generated, co-moving pH-field,[42] and the corresponding DO-flow along the substrate points towards the low pH region (green arrows). It accretes tracers towards the pen,[22,23,39] from where they are expelled to the back leaving a trail of enhanced ink-particle density. In principle, the same mechanism could be exploited to co-assemble other small objects, e.g, initiator, (macro-)monomers, and cross-linker molecules, which could later be used to fix line centers containing high tracer density by UV curing. Note that in all figures regarding experiments, the substrate is tilted such that IEX and tracers move from right to left.

Samples are studied either in dark field (Figure 1b, see also Figure S1b-d, Supporting Information) or bright field (Figure 1c and Movie S1, Supporting Information). In the stationary state depicted in Figure 1c, the line drawn at $v_{IEX}$ = 7.7 µm s$^{-1}$ first narrows and then broadens again. We identify a line focus (red arrow) separating an initial formation zone of length $l_f = v_{IEX} t_f$ (where $t_f \approx$ 60-70 s denotes the time elapsing between the passage of the IEX and the line focus) from a subsequent decay zone. To quantify the underlying mechanisms, we show the co-moving, height averaged pH-field recorded in bright field using 3-channel photometry in Figure 1d.[42,43] The arrows denote the local pH-gradient directions. We observe a stationary, elliptically distorted, and rather diffuse pH distribution. From this, we calculate the pH variation at the substrate height following Möller et al. (Figure 1e).[42] Note that at this height, the gradients driving the DO-flow are much more pronounced. Figure 1f shows the resulting trajectories of individual tracers relative to the IEX. Tracers in the line of the IEX motion approach from the front and leave at the back. Tracers to the sides show bent trajectories as they are swept towards the IEX. We depict the instantaneous lateral solvent flow field in Figure 1g in lab coordinates and with color-coded flow strength. Note the striking similarity with Figure 1e. Solvent flows towards the IEX but moreover also towards and along the lateral pH-minimum extending in the wake of the IEX. Due to its incompressibility, the solvent has to flow upward at the IEX.[32]

To explore the influence of the IEX speed on line writing, we varied the tilt angle $\vartheta$ to change $v_{IEX}$ (Figure S1b-d, Supporting Information). To good approximation, the IEX speed increases linearly with sin($\vartheta$). For IEX speeds in the range of 1.5 µm s$^{-1}$ < $v_{IEX}$ < 11 µm s$^{-1}$, straight single lines of long extension are obtained, with maximum contrast observed for 6 µm s$^{-1}$ < $v_{IEX}$ < 10 µm s$^{-1}$. At lower $v_{IEX}$, tracers become trapped in asymmetric assemblies close to the IEX. At larger $v_{IEX}$, lines get rather faint and occasionally a short-lived line split is observed, with the two lines merging by diffusion.

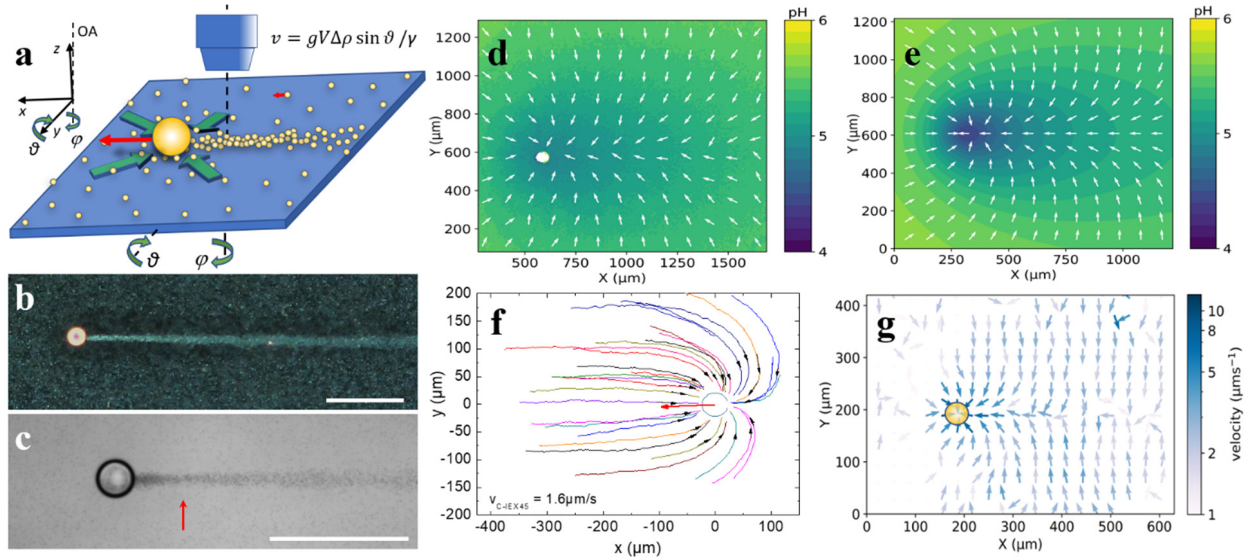

**Figure 1.** Writing of straight lines. a) Sketch of the experimental situation. The sample is placed on a stage allowing for tilting by $\vartheta$ and rotation around the optical axis (OA) by $\varphi$. The IEX (large sphere) rolls under gravity (large red arrow) through settled tracer particles (small spheres), rolling at much

smaller speed (short red arrow). Tracers are swept towards the IEX by the co-moving solvent flow field (green arrows) and assembled in the back of the IEX into a line of positive density contrast, which is left behind and disperses slowly by diffusion. b) Dark field image of C-IEX45 rolling on an inclined glass substrate ($\vartheta = 3.1°$, substrate $\zeta$-potential $\zeta_S = -105$ mV, $v_{IEX} = 7.7$ μm s$^{-1}$) in a suspension of Si832 at $c = 0.1$ wt.% (tracer $\zeta$-potential $\zeta_T = -68$ mV). Scale bar: 200 μm. c) The same but in bright field b/w and at $c = 0.02$ wt.%. The red arrow denotes the location of the line focus. d) Height averaged pH field in lab coordinates as determined by 3-channel micro-photometry for the experiment shown in panel (b). The pH-values are colour coded as indicated in the key. Arrows denote the local gradient direction. e) Ground pH-field as derived from panel (d). f) Map of tracer approach trajectories in IEX-relative coordinates for C-IEX45 rolling at $v_{IEX} = 1.6$ μm s$^{-1}$. The red arrow denotes the rolling direction. g) Tracer velocity field in lab coordinates as determined from Particle Image Velocimetry for the situation in panel (c). Local velocities are colour coded according to the key.

## 2.2. Line stability

For writing complex patterns, lines should be durable. The line in **Figure 2**a was written in Si832 at $c = 0.1$ wt.% for increasing times after IEX passage, which defines $t = 0$ (see also Movie S2, Supporting Information). Figure 2b shows the line profiles fitted by Gaussians, from which we derive the linewidth in terms of the Full Width at Half Maximum (FWHM). The FWHM increases slowly and at continually slowing pace with increasing time. Notably, even after some ten minutes, the linewidth has increased from 40 μm to merely 90 μm (Figure 2). The relative height of the maximum has decreased correspondingly by some 50%, and the line remains clearly visible (Figure 2a). We further studied the linewidth evolution for different $c$ and $v_{IEX}$. The linewidth increases with tracer concentration and decreases with IEX speed. The double logarithmic plot of Figure 2e shows the time-dependent FWHM scaled to $v_{IEX} = 8$ μm s$^{-1}$ and $c = 0.1$ wt.%. Data arrange on a single straight line of slope 0.48. A systematic analysis suggests an overall scaling of the line width as FWHM $\propto v^{-1} c^{1/2} t^{1/2}$ (Figure S2, Supporting Information). This scaling is compatible with accretion at constant solvent flux and purely diffusive line dispersion.

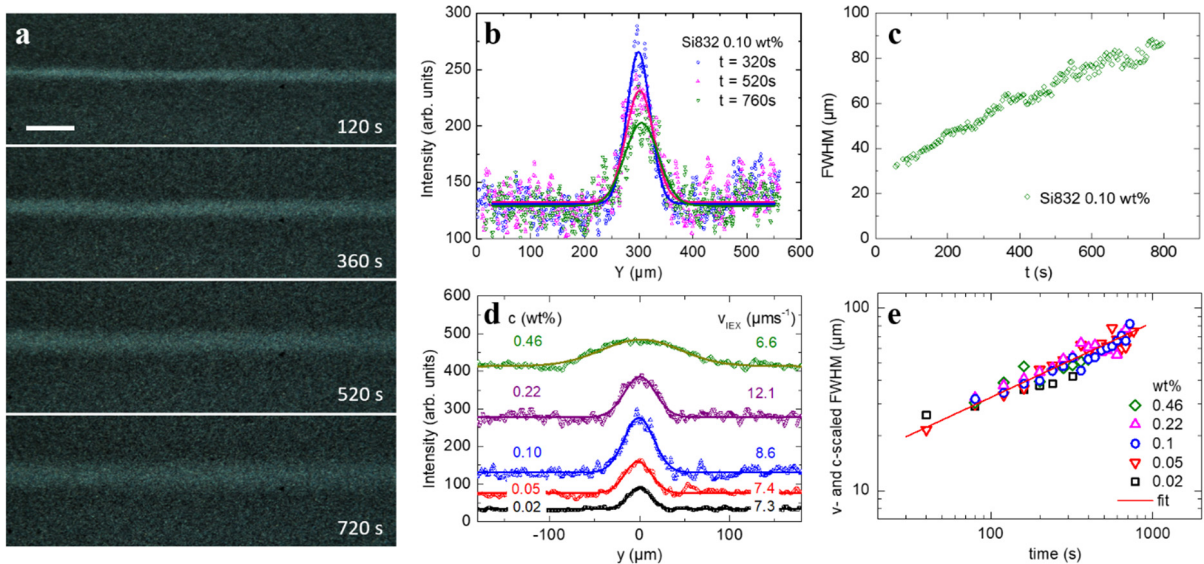

**Figure 2.** Line evolution past the formation zone. a) Dark field images of a line written by C-IEX45 in 0.1 wt.% Si832 for different times after IEX passage, as indicated. Scale bar: 100 μm. b) Line profiles (symbols) in lab coordinates for three selected times as indicated in the key. The solid lines are fits of Gaussians with an offset corresponding to the background scattering intensity. c) Full Width at Half Maximum (FWHM) as determined from fits to line profiles taken at different times $t$ after IEX passage. Data are taken after the passage of the line focus. The increase in linewidth slows with time. d) Line

profiles in IEX-relative coordinates obtained at different tracer concentrations and IEX speeds as indicated in the keys. These averages over 50 individual line profiles sampled over 100 μm along the lines were recorded at $t = 80$ s. All profiles are well described by fits of a Gaussian with an offset corresponding to the background scattering intensity. e) Scaling analysis: We separately determined the $c$-dependence and the $v$-dependence of the FWHM and scaled the obtained fit results to $c = 0.1$ wt.% and $v_{\text{IEX}} = 8$ μm s$^{-1}$. Data arrange on a straight line in this double logarithmic plot. From the fit of a linear function, we obtain FWHM $\propto t^\lambda$ with $\lambda = 0.48\pm0.04$, which is very close to the theoretically expected value of $\lambda = 0.5$.

### 2.3. Effective attraction

The line formation results as the net effect of the pH-driven DO-flow and typically involves Peclet numbers of Pe ≈ 10-30. Thus, the line formation is dominated by tracer advection caused by the DO-flow. The Reynold numbers, however, are small (Re ≈ 10$^{-4}$), i.e., the flow is dominated by viscous forces and thus, laminar. Furthermore, the ratio of the tracer mass and drag coefficients is small ($m_\text{T}/\gamma_\text{T} \approx 10^{-7}$ s), and therefore, the motion of the tracers is overdamped. Overall, the DO-flow induced by the IEX leads to a directed motion of the tracers towards the moving IEX, whereas the IEX does not significantly respond to the tracers. Thus, the effective interaction between the IEX and each tracer can be described as an effective non-reciprocal attraction exerted by the IEX on the tracers leading to a center of mass motion of the tracers.[33] We extracted the form of these effective interactions from experimentally measured velocities obtained from video-tracking (**Figure 3**a). We found that the tracer velocity field at a distance $r$ behind the IEX can be fitted by the following form:

$$|\vec{v}(r)| = \frac{c_1}{r^\alpha + c_2}. \tag{1}$$

Thus, it yields a non-reciprocal attractive effective force of strength $F_\text{a}(r) = \gamma_\text{T}|\vec{v}(r)|$ reaching out some hundreds of microns (inset in Figure 3a and Figure S11, Supporting Information). Here, $\gamma_\text{T}$ denotes the Stokes drag coefficient of the tracers. The parameters $c_1$, $c_2$, and $\alpha$ are obtained from a nonlinear least-squares fit (see Figure 3a and Supporting Information for details).

### 2.4. Model

Based on the effective force, we formulate a minimal (particle-based) model for the tracer dynamics at prescribed motion of the IEX. Within this model, the tracers are considered as overdamped Brownian particles, which experience a non-reciprocal effective force $\vec{F}_{eff}(t,\vec{r})$ due to the IEX as obtained from Equation (1) (see also inset of Figure 3a and Methods Section) and are subject to thermal diffusion. The position of the $i$-th tracer particle evolves in time according to

$$\dot{\vec{r}}_i(t) = \frac{1}{\gamma_\text{T}}\left[\vec{F}_{eff}(t,\vec{r}) - \sum_{j=1,j\neq i}^{N} \nabla_i u(|\vec{r}_i - \vec{r}_j|)\right] + \sqrt{2D}\vec{\xi}_i(t), \tag{2}$$

where $\gamma_\text{T}$ denotes the Stokes drag coefficient, $D = k_\text{B}T_\text{bath}/\gamma_\text{T}$ the diffusion coefficient with bath temperature $T_\text{bath}$, and $\vec{\xi}_i(t)$ denotes Gaussian white noise with zero mean and unit variance. The optionally considered (repulsive) pair interaction between the tracer particles is modelled by a Weeks-Chandler-Anderson potential $u(r)$.[44] For simplicity, we assume that the total effective force $\vec{F}_{eff}(t,\vec{r})$ is radially symmetric (see Methods Section for details). Clearly, this is a simplification of the overall DO-flow pattern (Figure 1g and Figure S5, Supporting Information), but, as we shall see, sufficient to capture the essentials of line writing. Motivated by the experimental observations (Figure 1), we further neglect the very slow collective tracer motion under gravity and any diffusio-phoretic motion of tracers originating from DO flows along the surfaces of the mobile tracers due to the concentration gradient.[22] The corresponding Langevin equations for the tracer dynamics [Equation (2)] are solved numerically with a deterministically moving IEX at prescribed velocity with and without considering direct tracer-tracer interactions $u(r)$ to check the effect of the latter on the line formation (see Methods Section and Supporting Information for details). Finally, the effective attraction leads to the formation of straight, narrow, and durable lines of increased tracer density (Figure 3b and Movie S3, Supporting Information).

To access larger length and time scales, we additionally formulate an equivalent continuum model, which is given by the two-dimensional Smoluchowski equation for the probability density $p(t,X,Y)$ of the (point-like) tracer particles [Methods Section Equation (5)]. We solved the Smoluchowski equation numerically in the coordinate frame of the IEX [Methods Section Equation (6)] for a deterministically moving IEX with prescribed velocity. As shown in Figure 3c, our continuum model also leads to the formation of a straight and narrow line of increased probability density.

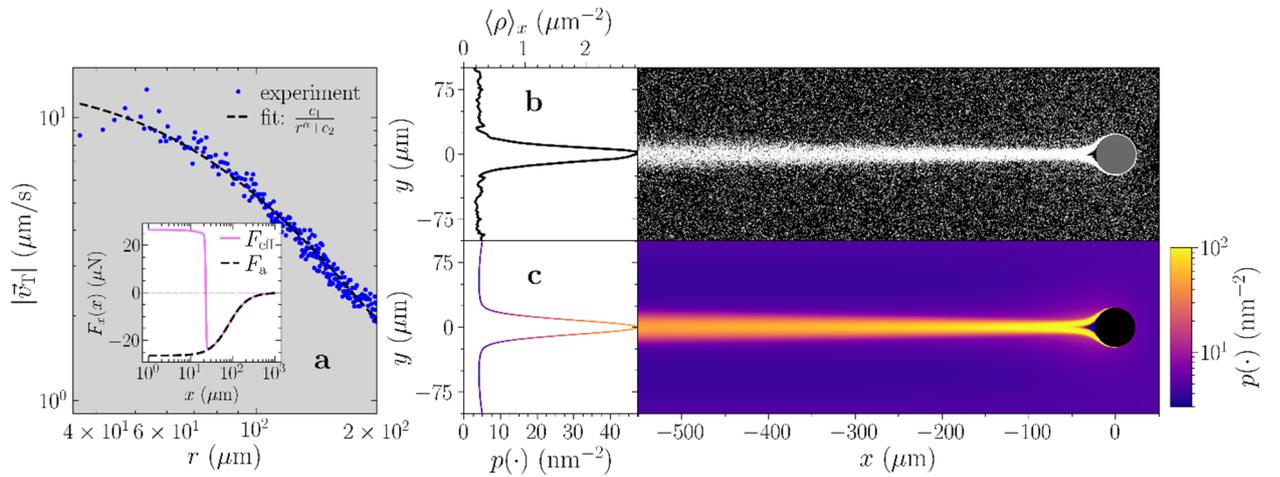

**Figure 3.** Modelling of line writing. a) Tracer approach speed averaged over an angular region of 30° behind the IEX rolling at $v_{IEX}$ = 1.6 µm s$^{-1}$ (blue) in dependence on the radial distance $r$ to the IEX center. The black dashed line is a nonlinear least-squares fit of Equation (1) returning $c_1 \approx (7.2 \pm 2.3) \times 10^4$ µm$^3$ s$^{-1}$, $c_2 \approx (5.3 \pm 1.9) \times 10^3$ µm$^2$, and $\alpha$ = 1.96 ± 0.06. Inset: $x$-component of effective forces as used in the continuum model. The attractive part $F_a$ corresponds to the fit in the main panel (black dashed curve), the full effective force $F_{eff}$ (solid pink curve) additionally accounts for volume exclusion interactions between the IEX and the tracers. b,c) Exemplary results from theoretical modelling based on the effective force shown in panel a in IEX-relative coordinates. Parameters were chosen to correspond closely to those of the experiments: $2a_{IEX}$ = 45 µm, $D$ = 0.5 µm$^2$ s$^{-1}$, $\gamma$ = 7.85 × 10$^{-9}$ kg s$^{-1}$, $v_{IEX}$ = 8.0 µm s$^{-1}$ (b) and $v_{IEX}$ = 8.6 µm s$^{-1}$ (c) (see also **Table 2**). b) Snapshot taken at $t$ = 100 s from a BD simulation starting from $N$ = 112,200 uniformly distributed point-like tracers. The line profile (left) is averaged over the interval $x \in$ [-550 µm, -500 µm]. c) Numerical solution of Equation (6) with initially uniform probability density $p_0$ = 5 × 10$^{-6}$ µm$^{-2}$ after $t$ = 2000 s. The probability density $p(\cdot)$ = $p(2000\ s, x, y)$ (right) and the line profile at $x$ = -550 µm (left) are color-coded as indicated in the key.

In the continuum model, the stationary state was typically reached within a few hundred seconds (Figure S12, Supporting Information). As in our experiments, the density distribution may take different shapes in the line formation region, but beyond the focal point, the line profile is Gaussian (Figure S14, Supporting Information). Note from the profile in Figure 3c that the lines are embedded within a region of marginally reduced tracer density reflecting the extension of the accretion zone. As in the experiments, this depleted accretion region serves to enhance the contrast of the main line. Our observations demonstrate that the main ingredients of line formation are simple and generic: Writing into water requires a quick, medium-range ink-transport towards the moving pen caused by an effective hydrodynamic attraction of ink particles towards the IEX and a slow, diffusive line decay. Thus, we term this novel type of writing a line of increased density in water *hydrodynamic writing* (HDW).

## 2.5. State diagram

Using the continuum model [based on Equation (6)], we systematically varied the IEX speed $v_{IEX}$ (equivalent to changing the tilt angle $\vartheta$ in the experiment) and the tracer diffusion constant $D$ (equivalent to changing the tracer particle size). To discriminate between different line types, we fitted a Gaussian or a sum of two Gaussians to the line profiles of the probability density taken in the stationary state after $t = 2000$ s at distance $x = -950$ μm behind the IEX (Figure S15, Supporting Information). We observed three different line types: single line, split line, and inverted line (**Figure 4**a). Consequently, the state diagram in the $v_{IEX}$-$D$-plane shows three distinct regions (Figure 4b).

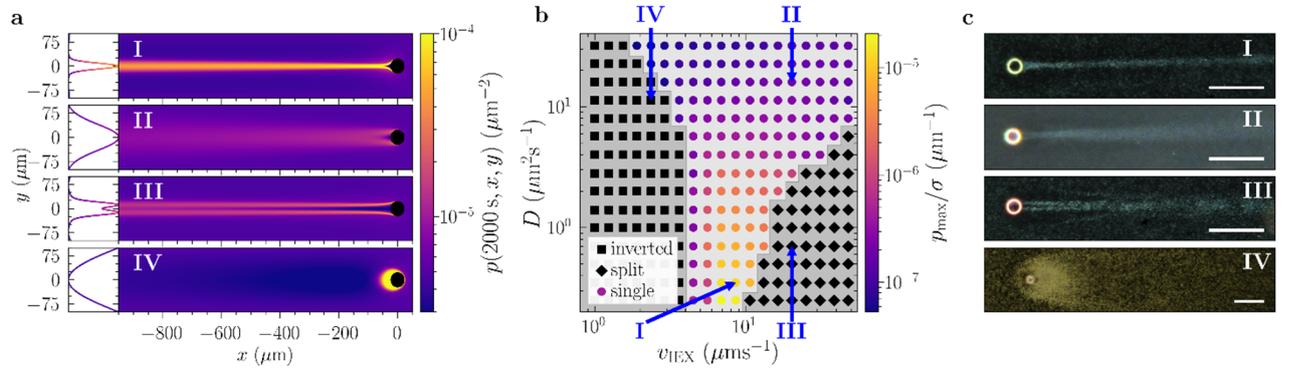

**Figure 4.** State diagram. a) Snapshots (right) and line profiles at $x = -950$ μm (left) obtained from the numerical solution of Equation (6) in the stationary state at $t = 2000$ s. The representative examples are denoted by Roman numerals: narrow single line (I), broad single line (II), split line (III), and inverted line (IV). Colours encode the probability density $p(2000\,s, x, y)$ as given in the key. b) State diagram in the $v_{IEX}$-$D$-plane (other parameters as before, see also Methods Section, Table 2). Symbols denote the three principal line types as discriminated by the line profiles at $x = -950$ μm: inverted line (squares), split line (diamonds), and single line (colour-filled circles). The state points corresponding to the examples shown in panel (a) are marked by blue arrows and Roman numerals. For single lines, colours encode the line sharpness $p_{max}/\sigma$, where $p_{max}$ denotes the amplitude of the fitted Gaussian line profile and $\sigma$ denotes its standard deviation. c) Representative experimental examples observed in the three regions of the state diagram: narrow single line written in 0.10 wt.% Si832 at $v_{IEX} = 9.2$ μm s$^{-1}$ (I), broad single line written with 0.24 wt.% Si444 at $v_{IEX} = 8.2$ μm s$^{-1}$ (II), split line written in 0.10 wt.% Si832 at $v_{IEX} = 12$ μm s$^{-1}$ (III), and a line in the region of inverted lines shown in panel (b), here written in 0.1 wt.% Si2.1 at very low $v_{IEX} \approx 0.2$ μm s$^{-1}$ (IV). Scale bars: 250 μm.

Pronounced single lines are obtained at intermediate velocities (colour coded region with state points I and II in Figure 4b). We define a line sharpness $p_{max}/\sigma$, where $p_{max}$ denotes the amplitude of the fitted Gaussian line profile and $\sigma$ denotes its standard deviation. Line sharpness, and thus, contrast, is largest at small $D$. With increasing diffusivity, sharpness decreases and lines get continuously broader and fainter. Towards larger velocities, the line formation length expands and results in a split line (III). The simulated profile at $x = -950$ μm now shows two maxima, which eventually merge at larger distances $x$ by diffusive broadening. The regime of split line formation is thus located in the lower right corner of the state diagram (low $D$, large $v_{IEX}$).

Interestingly, the model further reveals a third line type at low velocities. Due to the low IEX velocity, a substantial amount of attracted tracers is trapped and no line of positive density contrast is written. Instead, in the steady state, a stationary cloud of tracers forms in the vicinity of the IEX due to a balance between trapping and diffusion, which we also observed in our experiments (example IV in Figure 4c). Accretion of tracers from the IEX surroundings leads to a broad shallow minimum in the line profile, i.e., the line contrast is inverted (see example IV in Figure 4a and Figure S15a, Supporting Information). The width of the minimum mirrors the $y$-extension of the range of attraction. Note that such a shallow depletion zone is also present at larger IEX velocities in the single-line regime (HDW), but as in the experiments, it is masked by a much more pronounced HDW line and increases the line contrast (Figures S1c and S15b, Supporting Information).

Overall, also lines written in the model vary considerably in quality but not all appear to be useful for actual writing. In practice, all inverted lines written at low $v_{IEX}$ remain too faint. Additionally, for increasing velocities, the flux of incoming tracers is reduced and becomes too small for providing a large density contrast for the inverted line profile. Within all regimes, high diffusivities lead to a significant decrease of the line sharpness resulting again in lines unsuitable for actual writing. However, our model clearly predicts that single narrow lines of large density contrast are obtained for low diffusivity and intermediate IEX speed, exactly as observed in the experiments (cf. the orange/yellow region in Figure 4b as compared to Figure 1b,c and Figure S1b-d, Supporting Information).

### 2.6. Graphical application

Having explored the experimental and theoretical aspects of line writing in some detail, we turn to its graphical application. Here, we again use weakly charged tracers and moderate writing velocities, which is essential for proceeding beyond straight lines by combining line writing with steering. In the experiments, we controlled line orientation using a manually programmable, motorized stage, allowing sequences of changes in tilt direction and angle. In the BD simulations we simply prescribed IEX speeds and trajectories. **Figure 5** shows representative examples, in which we successfully produced some simple patterns and symbols (Figure 5a-f), drew complex patterns with multiple line crossings (Figure 5g) or wrote text (Figure 5h,i). Next, we discuss some practical issues encountered during their production. Downscaling is addressed for the example of a pattern with rectangular turns in Figure 5a-c. The right angles and straight intervals start washing out when the length of straights approaches three to two times the line thickness. Another issue is line-shifting. In the simulation run of Figure 5d, we compare the written line to the prescribed sinusoidal IEX trajectory. Due to the relatively long-ranged tracer attraction (Figure 3a) and the continuing DO-flows towards the back of the IEX (Figure 1g), the written line is dragged along in the overall propulsion direction. This results in a shift of the final line as compared to the prescribed IEX trajectory. The intended sinusoidal shape, however, is fully retained. This is different for the case of line crossing. For instance, in the simulation run of Figure 5e, the initial upward stroke of the lower-case Greek letter phi is bent to the left (see also Figure S18, Supporting Information). In addition, the freshly written line shows a local enhancement of line thickness (blob) also shifting in the propagation direction of the IEX. By contrast to blobbing by transient sticking of the IEX to the substrate (Figure S20, Supporting Information), blobbing upon line crossing is systematic. However, while it is kept at bay by high writing velocities ($v_{IEX} \geq 10$ μm s$^{-1}$), line bending is not. The inset of Figure 5e shows a line crossing experiment at large $v_{IEX}$. Here, the duration of tracer attraction from the crossed high-density region is reduced, but the DO-flows are retained, and the line is still bent. In principle, line crossing can be avoided using multiple pens on merely touching trajectories (Figure 5f). Where it could not be avoided, we performed a stepwise adjustment of the stage tilting sequence to compensate for the observed and anticipated deviations from the desired line path (Figure S19, Supporting Information). With some experience, even complex patterns with multiple line crossings under desired angles, negligible line-bending, and little blobbing can be reproducibly drawn (Figure 5g). Note that the base of the "house" shown in Figure 5g is merely 500 µm and the height is about 850 µm, i.e., the house´s size is on the order of a single letter in this text. The figure was drawn within about five minutes at an IEX velocity of approximately 15 µm s$^{-1}$. It retained its characteristic shape for more than 15 min. Writing of individual letters is less difficult. Picking up a certain tradition,[9,10,12,14,15,20] we display the affiliation initials of the experimental and theoretical group in Figure 5h and i, respectively.

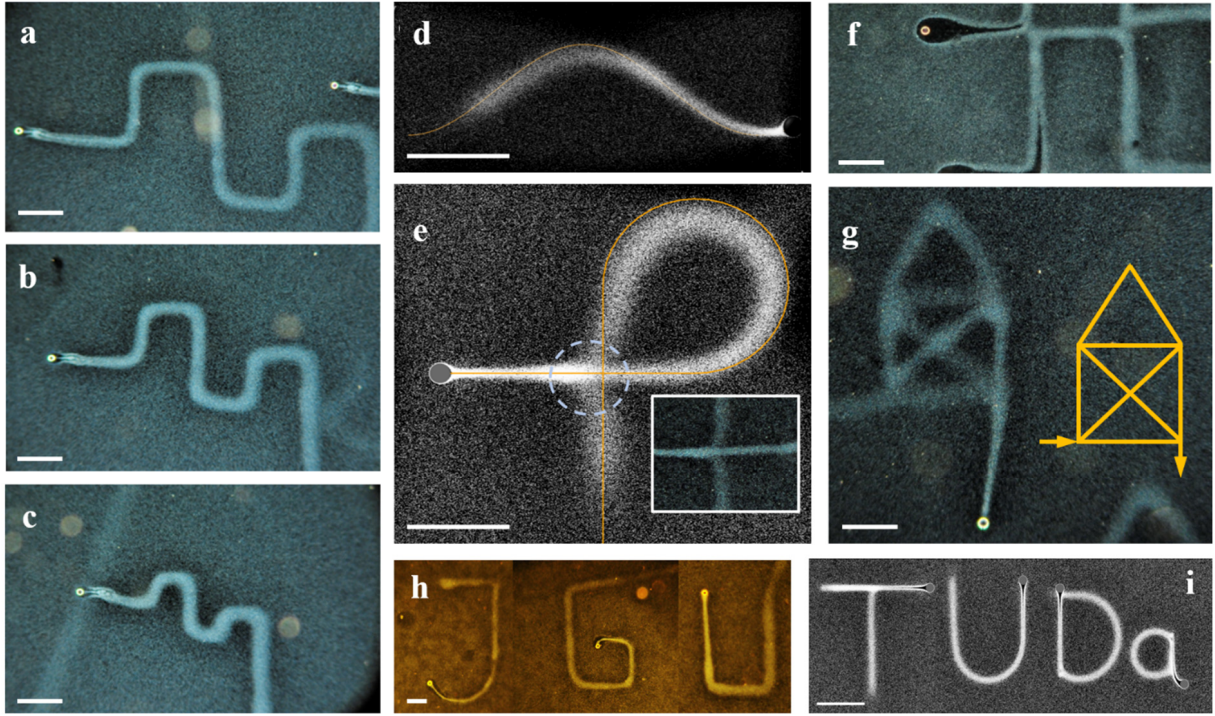

**Figure 5.** Examples of writing and drawing (experiments with C-IEX45 in Si832 and BD simulations with parameters as given in Table 2). All scale bars: 250 µm. a-c) Successive downscaling of pattern size. The length of the horizontals is 500 µm (a), 350 µm (b), and 175 µm (c). d) Sinusoidal line written in the BD simulation with parameters close to typical experimental ones (see Methods Section, Table 2). Note the slight, leftward shift of the written line with respect to the IEX trajectory (solid yellow line). e) Lower case Greek letter phi with loop and line crossing. Note the bending of the crossed line and the formation of a region of enhanced density along the freshly written line (dashed circle). Inset: Writing at high IEX speed of $v_{IEX}$ = 10 µm s$^{-1}$ reduces blobbing but retains line bending. f) Avoiding blob formation upon line crossing by mere line touching. g) One-line drawing of a complex pattern with rectangular and 45° angles as well as multiple line crossings. The inset shows the intended result ("Das Haus vom Nikolaus"). h) Initials of the Johannes Gutenberg University, Mainz, written with C-IEX45 in 0.2 wt.% Si832. i) initials of the Technical University of Darmstadt written in a BD simulation with $v_{IEX}$ = 12 µm s$^{-1}$.

## 2.7. Graphical fine tuning

So far, we concentrated on writing of and with durable, narrow lines of enhanced tracer density (positive contrast). This was facilitated by rolling C-IEX45 at speeds of $v_{IEX}$ = 6-10 µms$^{-1}$ through weakly charged Si832 ($\zeta_T$ = -68 mV) at 0.1-0.15 wt.% on native substrates ($\zeta_S$ = -105 mV). One may, however, wish to fine-tune the graphical appearance of the written lines. To this end, we exploit the modularity of our approach allowing tests of alternative inks, pens, and pen-drives. These additional and complementary experiments are described in detail in the Supporting Information (Figures S3-10). In short, tracers of larger or smaller size can be used to alter the line graininess and contrast. Replacing weakly charged tracers by highly charged tracers introduces additional diffusio-phoretic tracer motion. This alters the shape of the formation zone but leaves the decay zone unaffected (see also Movie S6, Supporting Information). The driving pH gradient can be manipulated by using different pen-types. Removing the pH gradient using chemically inert pens produces poor lines, while doubling the pH gradient using a pair of IEX yields thicker lines with weakly charged tracers and split lines with highly charged tracers. Replacing the anionic by a cationic IEX switches the sign of the pH gradient. Tracers are then repelled and the IEX carves a tracer free line into the background (negative contrast). Finally, we also tested different pen-driving mechanisms such as modular microswimmers self-assembling on horizontal substrates. These microswimmers propel autonomously and steer by more or less random rearrangements of the assembled tracers. Lacking a directed steering, microswimmers write curved lines of otherwise good quality (see also Movie S7, Supporting Information).

## 3. Conclusion

The present approach of writing by assembly features several advantages over other approaches. As ink is available all along the trajectory, writing by assembly eliminates the necessity of a large (potentially eddy-creating) reservoir required in deposition-approaches as well as any re-filling issues. Working with small pens in a viscous fluid at low Reynolds numbers for the pen motion ($Re_{pen} \approx 10^{-4}$-$10^{-3}$) avoids eddy-creation, leaves the fluid practically undisturbed, and allows writing of narrow lines of a few tens of µm thickness, which compares favourably to the line widths reached in microprinting. With this approach, ink was collected from a region of a few pen diameters in width at Péclet numbers on the order of $Pe_{ink} \approx 5$-$50$. This was large enough for a quick build-up of significant density modulations. Chosen pen-speeds of $v_{pen} > 1.5$ µm s$^{-1}$ were large enough to avoid binding of the assembled micro-sized tracers to the pen and thus lead to line formation by tracer expulsion. At the same time, $Pe_{ink}$ was small enough to avoid excessive direct transfer of kinetic energy to the ink particles. (This so-called dynamical friction may spoil the formation of highly visible lines, e.g., in the wake of a large star crossing a background of small stars in a globular cluster.[45,46]) As a result, line dispersion occurred by diffusion only.

We demonstrated a facile and inexpensive approach to write on the micron scale within a liquid medium. A variety of further technical extensions awaits realization. Parallel writing of finely structured, large-scale density patterns becomes possible with several pens. Interruptions between words as well as dashed lines could be facilitated using photo-switchable chemical sources. Guided motion of the pen can further be realized by mechanical guiding,[47] optical[48] or magnetic forces,[49] and chemical fields.[50] Exploiting steering by chemical gradients would allow visualizing faint chemical traces left by other objects. Additionally, steering pH sources with optical tweezers in 3D buoyant ink dispersions would open access to freely suspended 3D patterns of arbitrary shape. Beyond global erasing by sonication, typo correction and line reconfiguration may be implemented by rewriting with or without prior heating with an IR laser (see also Movie S8 and S9, Supporting Information). Conversely, one could co-assemble suitable chemicals for line fixing like a mixture of photo-initiators, monomers, and cross-linker molecules. The important objective of further downscaling into the few-µm-range appears to be feasible for fluorescent tracers imaged by fluorescence microscopy and assembled by thermophoresis at a micron-sized heat source steered by laser tweezing. Adaptive learning and supervised machine learning could be used to program trajectories leading to the analogue of handwriting. Already these few examples of possible future developments highlight an interesting point on the technical side: namely the versatility of our generic approach of combining ink assembly via long-ranged attractions with purely diffusive decay and programmable steering of the pen. It comprises a high degree of modularity, allows for the exchange of the constituents (pen, ink, drive) as to one´s requirements, and is open for the combination with further modules. Thus, we anticipate that our results and conclusions are valid for a broad class of systems and may open routes to many future enterprises including the structuring of fluids, the visualization of chemical traces, the assembling of functional objects like actuators or drug carriers, information storage, and numerous artistical applications. Finally, our results could also be used to prepare desired initial states for future colloid experiments, e.g., on collective diffusion. Drawing fine lines, durable patterns, and individual letters into water was only the first step.

## 4. Methods Section

*Sample preparation*: As pens, we mainly used micro-gel based cationic IEX beads (C-IEX45, Purolite Ltd, UK). The mean diameter ($2a = 45 \pm 2.3$ μm) was measured by image analysis. The IEX beads were recharged with $H^+$ ions by washing them twice in aqueous HCl (20 v/v%) solutions and subsequently rinsing them with deionized water. The beads were then stored for 48h in an exicator. For some experiments, we also used anionic resin-based IEX beads or splinters thereof (A-IEX15, Mitsubishi Chemical Corporation, Japan and A-IEX-L, Carl Roth + Co. KG. Karlsruhe, Germany). These were only rinsed with deionized water and subsequently dried.

Passive ink particles were sulphate-stabilized polystyrene (PS) or silica (Si) colloidal suspensions (Microparticles, GmbH, Germany). **Table 1** denotes the lab codes, Manufacturer batch nos., sizes, and ζ-potentials where applicable. Prior to use, a part of the original stock (32 wt.%) was diluted with deionized water to 1 wt.% and conditioned for some weeks in a flask containing mixed-bed IEX (Amberjet, Carl Roth GmbH + Co. KG, Karlsruhe, Germany) to remove excess ions from the production process. After deionization, the IEX was removed, and the deionized suspensions stored in contact with ambient air. Thus conditioned, they show comparably low ζ-potential magnitudes due to $CO_2$ adsorption.[51] $CO_2$ absorption can be reversed and any absorbed $CO_2$ removed by further contact with mixed bed IEX in strictly gas-tight flasks.[51] As compared to the merely deionized state, the deionized and de-carbonized state shows an approximately doubled ζ-potential magnitude.

The sample cell was constructed by sandwiching a chemically inert PVC ring (inner radius R = 12.5 mm, height H ~ 0.5 mm) between two standard microscopy slides (75×25×1mm, soda lime glass, hydrolytic class 3, VWR International, Germany). The ring was fixed to the bottom slide with epoxy glue (UHU plus sofortfest, UHU GmbH, Germany). Prior to cell assembly, the glass slides were thoroughly cleaned to remove coatings and organic contaminants (washing in 1 v/v% Hellmanex®III-solution [Hellma Analytics, Germany], sonication in isopropanol for 30 min., rinsing with deionized water, and drying in $N_2$ flow). Right after cleaning, their ζ-potential was ζ = -138 ± 8 mV. If stored in the dry state under ambient conditions, the ζ-potential magnitude decreases and equilibrates at ζ = -105 ± 5 mV. The magnitude can be reduced further to ζ = -70 ± 5 mV by immersing the slides for one hour in diluted deconex® cleaning solutions (5 v/v%, deconex®11, VWR, Germany).

**Table 1: Tracer colloids and IEX beads.** Parameters of tracer colloids and IEX beads used in this work. ζ-potentials were determined from electrokinetic experiments using a custom-built Doppler Velocimeter.[52]

| Lab Code | Manufacturer Batch number | Diameter, 2$a$ (µm) | ζ-potential under deionized conditions (mV) |
|---|---|---|---|
| Si444 | SiO2-F-SC54 | 0.442 ± 0.054 | -70 ± 3 (ambient $CO_2$) |
| Si832 | SiO2-F-L1287 | 0.839 ± 0.031 | -68 ± 3 (ambient $CO_2$) <br> -108 ± 5 (de-carbonized) |
| Si2.1 | SiO2-F-L4255-1 | 2.11 ± 0.18 | N/A |
| PS2.3 | PS/Q-F-L2090 | 2.30 ± 0.22 | -35 ± 3 (ambient $CO_2$) <br> -65 ± 5 (de-carbonized) |
| C-IEX45 | CGC50 x 8 | 45 ± 2 | N/A |
| A-IEX-L | K306.2, Amberlite | 250-400 | N/A |
| A-IEX15 | CA08S | 15.0 ± 1,6 | N/A |

*Line imaging*: Cells were mounted on the stage of an inverted microscope (DMIRBE, Leica, Germany) and writing was observed at 10x magnification with a 12.87 Mpx. consumer DSLR (D700, Nikon, Japan) using a 0.63x mounting tube. In the images, 100 px correspond to 0.93 µm. Images were captured in 14bit RAW format at intervals of 4s, converted to TIFF, and stored. Videos and cropped images for display were constructed from the TIFF colour-images. Data for further evaluation were converted to grey-scale and analysed using a custom-written Python script. For $c \leq 0.2$ wt.% of Si832, the recorded intensity was proportional to the tracer concentration, showing the absence of multiple scattering effects (Figure S2c, Supporting Information). Line profiles are constructed from intensity readings perpendicular to the line axes. For an individual profile, each $I(y)$ is an $x$-average over ten neighbouring pixels. Typically, 50 individual profiles obtained from locations distributed over a distance of ~100 µm along the written line were averaged to improve statistics. From fits of a Gaussian to the averaged profile, we extracted the line amplitude, standard deviation, and full width at half maximum.

*Particle image velocimetry*: For flow-field imaging, we studied the stationary-state motion of tracers using either the DSLR at 10x magnification and 20 fps or (for smaller tracers) a fast monochromatic camera (acA133-200um, Semi Python 1300, Basler, Germany) at 20x magnification and 100 fps. Flow fields were calculated from image pairs separated by $\Delta t = 100$ ms using the OpenPIV Python package. For each location, the results were averaged over 50 successive image pairs.

*pH microphotometry*: For pH mapping, we employ a further refined version[42] of the general photometric approach reported by Niu et al.[43] A diluted mixture of two universal indicators (pH 4-10, Merck, Germany; pH 1-5, Sigma-Aldrich, USA, ratio 1:7) was injected into the sample dispersion ($c = 100–200$ µmol L$^{-1}$). Images were recorded in transmission under Köhler illumination ($\Delta I(x,y)/\langle I \rangle = 0.015$) using the DSLR camera. The 4.256×2.832 pixels on the CMOS-sensor are arranged in a Bayer pattern for the three RGB-channels resulting in dots of four px each, defining the maximum spatial resolution. For each dot, we compared the recorded channel intensities to those of a reference sample (deionized water) to obtain the absorbance in each channel from the Beer-Lambert-Law. Calibration was performed using pH buffer solutions ranging from pH 1.9 to 8.9 in 0.5 pH steps. For intermediate values, we used bicubic interpolation. We improved the signal-to-noise ratio by binning to 6x6 dots, reducing the final resolution to 357x237 blocks. At 10x magnification, we have 5.8x5.8 µm² per block. In the pH-range of interest (4 to 6), a height averaged pH-gradient resolution of 0.02 pH steps/6 µm results. Results are modelled by numerical solutions of the three-dimensional advection–diffusion equation, which also yields the pH-maps at specific locations, e.g., at the cell bottom.[42]

*Theoretical model*: The tracer particles are modelled as overdamped Brownian particles with radius $a_T$ and are subject to an effective external force $\vec{F}_{eff}(t,\vec{r})$, which is centred around the IEX position and is explicitly time-dependent due to the IEX motion. The position of the *i*-th tracer particle evolves in time according to Equation (2). The effective (non-reciprocal) force $\vec{F}_{eff}(t,\vec{r})$ describes the net impact of the IEX on the dynamics of the tracers. It is given as a sum of a short-ranged repulsion and a long-ranged attractive force $\vec{F}_a$ [Equation (1)]. The former accounts for steric repulsion between tracers and IEX. The latter accounts for the net effect of the hydrodynamic flow that is created by the IEX. We found empirically (Figure 3) that in the back of the IEX, the flow field decays with the distance $\vec{R}(t) := \vec{r} - \vec{r}_{IEX}(t)$ from the IEX located at $\vec{r}_{IEX}(t)$ and moving deterministically at constant speed $v_{IEX}$ as

$$\vec{F}_a(t,\vec{r}) = -\frac{\gamma_T c_1}{\left(|\vec{R}(t)|^\alpha + c_2\right)} \frac{\vec{R}(t)}{|\vec{R}(t)|}. \tag{3}$$

The constants $c_1$ and $c_2$ and the exponent $\alpha$ are determined from a fit to the velocity measurements (Figure 3 and Figure S11, Supporting Information). The uncertainties in $c_1$ and $c_2$ are rather large due to the interference of upward advection with lateral attraction for $|\vec{R}(t)|$ < 60 µm. For simplicity, we further neglect the asymmetry of the flow field. We compensate for both by introducing an ad hoc factor 2 using $c_1 \approx 3.6 \times 10^4$ µm³ s⁻¹. This results in a quantitative match of experimental observations and model predictions over an extended parameter range (Figure 4).

*Brownian dynamics simulations*: The equations of motions for the tracer particles were solved numerically in two spatial dimensions on an area $A = L_x L_y$ with periodic boundary conditions and area fraction $\varphi = N\pi a_T^2/A$ by using LAMMPS.[53,54] The repulsive part of the effective IEX-tracer interaction is modelled by a shifted Yukawa potential with strength $\varepsilon_{Yuk} = 10 k_B T_{bath}$, range $\sigma_{Yuk} = 1$ µm, and shift $\Delta r = a_{IEX}$.[55] The time steps are $\Delta t = 10^{-4}$ s for non-interacting tracer particles and $\Delta t = 10^{-5}$ s for repulsively interacting tracer particles. All other simulation parameters are summarized in **Table 2**.

**Table 2:** Simulation parameters. Default parameters used for the BD simulations and the numerical solution of the continuum model if not stated explicitly.

| Parameter: | Value: |
|---|---|
| $a_{IEX}$ | 22.5 μm |
| $a_T$ | 0.412 μm |
| $\gamma_T$ | 7.854 × 10$^{-9}$ kg s$^{-1}$ |
| $v_{IEX}$ | 8.0 μm s$^{-1}$ |
| $D$ | 0.5 μm$^2$ s$^{-1}$ |
| $\varphi$ | 0.152 |
| $T_{bath}$ | 298 K |

*Continuum theory*: To describe the density distribution of the tracer particles in presence of the moving IEX, we formulate the Smoluchowski equation, which describes the time evolution of the probability density for finding a tracer particle at a point $(t,X,Y)$ in space and time, where $X$ and $Y$ denote the coordinates in the laboratory frame.[56] For simplicity, we neglect any direct tracer-tracer interactions here and set $u(r) = 0$ in Equation (2). We only account for the net effect of the flow field and for the volume exclusion between the IEX and the tracers using the total effective force

$$\vec{F}_{eff}(t,\vec{r}) = \vec{F}_a(t,\vec{r}) \tanh\left(\sqrt{(X - v_{IEX}t)^2 + Y^2} - a_{IEX}\right), \quad (4)$$

with $\vec{r} = (X, Y)$ and IEX-radius $a_{IEX}$ (see inset of Figure 3a). The IEX moves along the $x$-axis at a constant speed $v_{IEX} > 0$. We transformed the corresponding Smoluchowski equation for the probability density $p(t,X,Y)$ (we omit the arguments for clarity)

$$\frac{\partial p}{\partial t} = -\frac{1}{\gamma_T} \nabla \cdot (\vec{F}_{eff} p) + D \nabla^2 p \quad (5)$$

into the coordinate frame of the IEX by applying the Galilei transformation $x \coloneqq X - v_{IEX}t$, $y \coloneqq Y$. The transformed equation reads

$$\frac{\partial \tilde{p}}{\partial t} = -\frac{1}{\gamma_T} \nabla \cdot (\vec{F}_{eff} \tilde{p}) + v_{IEX} \frac{\partial \tilde{p}}{\partial x} + D \left(\frac{\partial^2 \tilde{p}}{\partial x^2} + \frac{\partial^2 \tilde{p}}{\partial y^2}\right), \quad (6)$$

where now $\tilde{p} \coloneqq p(t, x, y)$. We omit the tilde throughout this work. Again, $D = k_B T_{bath}/\gamma_T$ denotes the tracer diffusion coefficient.

*Numerical solution method*: We solve Equation (6) numerically in two spatial dimensions on a rectangular area of size $(L_x, L_y) = (1000$ μm, $200$ μm$)$ with $x \in$ [-950 μm, 50 μm] and $y \in$ [-100 μm, 100 μm]. The initial condition is given by a uniform distribution $p(0, x, y) = 1/(L_x L_y)$. We use Dirichlet boundary conditions at the right, top, and bottom boundary with $p(t, 50$ μm, $y) = p(t, x, \pm 100$ μm$) = 1/(L_x L_y)$ and no-flux boundary conditions at the left boundary, i.e., $\partial_x p(t, x, y)|_{x=-950\,\mu m} = 0$. Finally, the Smoluchowski equation is solved with the NDSolve method from Mathematica[57] by using the method of lines,[58,59] finite element discretization,[60] and the parameters displayed in Table 2.

## Supporting Information

Supporting Information is available from the Wiley Online Library or from the authors.

## Acknowledgements

We thank A. P. Lanz for designing and realizing the programmable stage for writing as well as D. Botin and P. Vogel for the characterization of tracers and substrate surfaces. This work was supported by the Max Planck Graduate Center with the Johannes Gutenberg-Universität Mainz (MPGC). Financial Support of the DFG in the priority program SPP 1726 (Grant Nos. Pa 459/18-1 and Pa 459/18-2) is gratefully acknowledged. NM is fellow of the MPGC. LH gratefully acknowledges the support by the German Academic Scholarship Foundation (Studienstiftung des deutschen Volkes).

## Conflict of Interest

The authors declare no conflict of interest.

## Author Contributions

N.M. and L.H. contributed equally to this work. NM and TP developed the concept of writing in water. NM built, performed, and evaluated the experiments. RN contributed the single particle tracking experiments. LH and BL developed and performed the theoretical modelling. All authors contributed equally to data discussion and writing of the manuscript.

## Data Availability Statement

All data needed to evaluate the conclusions in the paper are present in the paper and/or the Supporting Information. Raw data are available upon reasonable request from the authors.

# Supporting Information

## Writing Into Water

*Nadir Möller, Lukas Hecht, Ran Niu, Benno Liebchen\*, Thomas Palberg\**

**This part includes:**

> Supporting Text
> Figures S1 to S20
> Captions of Movies S1 to S9

**Other Supporting Information for this manuscript include the following:**

> Movies S1 to S9 (see SMALL Website)

**Supporting Text**

**1. Additional Experimental Data**

The modular character of our approach allows for a systematic test of line properties upon changing the experimental boundary conditions, the employed components, and the pen-drive. We first present our experiments performed for the C-IEX45/Si832 system, then turn to other combinations.

**1.1. Variation of IEX speed**

On an inclined substrate, ion-exchange resin bead (IEX) and tracers roll downward. Their speed is easily tuned by varying the substrate tilt angle $\vartheta$. This is exemplarily shown in Figure S1a for C-IEX45 beads rolling in tracer-free water. Data are averaged over five beads. Error bars denote the statistical error in $v_{IEX}$, which increases with $\vartheta$, and the systematic uncertainty in meeting the adjusted $\vartheta$, which decreases with increasing $\vartheta$.

Single lines are written in a speed range of 1.5 µm s$^{-1}$ < $v_{IEX}$ < 10 µm s$^{-1}$. At lower speeds, we observe the formation of an extended, asymmetric accretion zone immediately past the IEX (Figure 4c example IV, main text). Surplus tracers leave this region through a wide contour and no useful line is obtained. At larger speeds, the line occasionally shows an initial split into two parallel lines separated by only a small distance. These lines quickly merge by diffusion. For concentrations $c \geq 0.2$ wt.%, larger velocities are needed to generate the line split. We conclude that moderate speeds return the best results.

**1.2. Line evolution past the line focus**

The line width in the decay region depends on both the tracer concentration and the IEX speed. Figure S2a shows snapshots of C-IEX rolling through Si832. From top to bottom, the tracer concentration $c$ increases, and the velocities vary in the range of $v_{IEX}$ = 6-12 μm s$^{-1}$. With increasing $c$, the background intensity increases and the lines appear thicker.

For each line in Figure S2a, we recorded the intensity profiles 80 s after IEX passage, i.e., far past the line focus and show these in Figure S2b. For better statistics, we here averaged over a trail length of 50 μm. All five density profiles are well described by Gaussians superimposed on a flat background. The fits return the standard deviation $\sigma$, which relates to the Full Width at Half Maximum (FWHM) as 2.355$\sigma$ = FWHM. From independent measurements on Si832 layers equilibrated at $\vartheta$ = 0 and performed under exactly identical illumination, we found the dark field scattered intensity to increase with a power law dependence as $I \propto c^a$ (Figure S2c). Fitting a linear function to the data in the double logarithmic plot of Figure S2b returns an exponent of $a$ = 0.85±0.04. This is close to but below the expectation of light scattering theory for dilute suspensions: $a$ = 1. The observed small deviation is attributed to tracer-tracer repulsion, which leads to the development of fluid order and in turn, decreases the forward-scattering. Figure S2d shows the time-dependent FWHM extracted from the concentration-dependent experiments in Figure S2a. The black arrow marks $t$ = 80 s used in Figure S2b. All curves increase roughly linearly in this double-logarithmic plot. The feature marked by the blue arrow is caused by the passage of a blob also seen in Figure S4a. As expected, the curves get wider for increasing concentrations, but the vertical spacing between the curves does not match exactly with the increase in weight fraction. This could be due to the different velocities of the IEX in Figure S2a. We therefore determined the velocity dependence of line widths in independent measurements at constant tracer concentration.

Figure S2e shows the data for $c$ = 0.18 wt.%. Here, we averaged over five to ten IEX beads per adjusted tilt angle. The large standard deviation in the measured IEX speed is attributed to the size dispersity of the IEX. The widths decrease systematically with increasing velocity. Data in this double logarithmic plot are well described by a power law with exponent -1.02 ± 0.09. This is remarkably close to -1. The comparably large uncertainty of the fit and the small deviation are again attributed to variations in IEX size. We use the result to rescale the data of Figure S2d to $v_{IEX}$ = 8 μm s$^{-1}$. The velocity-scaled data sets are displayed in Figure S2f. The data sets are now spaced more evenly. The velocity-scaled FWHM at $t$ = 80 s is shown in dependence on concentration in Figure S2g. The data in this double-logarithmic plot are well described by a power law of exponent $a$ = 0.51 ± 0.02. We use the result to rescale the data in Figure S2f to $c$ = 0.1 wt.%. The velocity- and concentration-scaled data are plotted double-logarithmically in Figure S2h versus time after IEX passage. All data arrange neatly along a single curve. Fitting a linear function to the data returns a power-law behavior FWHM $\propto t^\lambda$ with exponent $\lambda$ = 0.48 ± 0.04. This is close to the theoretical expectation of $\lambda$ = 0.5 for free diffusion.

We stress that these data were taken by averaging over only a few IEX at each inspected set of boundary conditions. The observed scaling behavior should therefore be taken with due caution. However, even these preliminary data are well compatible with a simple scaling behavior for the line width as FWHM $\propto v_{IEX}^{-1} c^{1/2} t^{1/2}$. The $v$-scaling would be expected as long as the IEX accretes the tracers with a constant solvent flux arriving at its surface. This is reasonable due to the large exchange capacity of the IEX and the low concentration of residual cations. The $t$-scaling is expected for a purely diffusive decay.

### 1.3. Shaping the formation zone

In this section, we consider the influences of tracer charge, substrate charge, and pH-gradient strength. For C-IEX45 rolling in Si832, we observe significant changes of the shape of the formation zone but no effects for the decay zone. The changes in the formation zone relate to a systematic variation in the approach-distance of the tracers.

The pH-gradient induces significant diffusio-osmotic (DO) flows along a highly charged substrate, which transport low-charge tracers towards the IEX. The approach velocity measured between 200 μm and 50 μm distance from the IEX center was used to describe this approach in terms of an effective attraction exerted by the IEX (Figure 3, main text). At shorter distances, the hydrodynamic flow pattern is significantly more complicated due to upward components in the solvent motion, the complex geometry of the wedge between IEX and substrate, hydrodynamic and direct tracer-tracer interactions, and the presence of an additional type of phoretic flow. In fact, low-charge or uncharged tracers are halted by these effects very close to or even underneath the IEX center. The exact stopping distance is not accessible in the flow field measurements due to shading by the IEX. We accounted for these phenomena in an effective way by adding a repulsive component to the total effective force.

The situation differs for highly charged tracers due to the additional diffusio-phoretic slip existing at their surface.[1,2] For this case, the approach situation is sketched in Figure S3. Relative to the tracer surface, the pH-gradient induces an additional inward solvent flow (DPS, dark blue arrow), which results in an outward tracer motion (DPT, violet arrow). In the sketched situation, the tracer would still be swept further inward towards the IEX. However, with increasing gradient strength, $v_{DPT}$ will increase until it equals $v_{DO}$, and the tracer becomes stalled.

To vary the relative weight of the two phoretic flows, we systematically varied the charge ratio $\chi = \zeta_T/\zeta_S$ between tracers and substrate. Experiments in the main text were performed with deionized tracer suspensions equilibrated in contact with ambient air. Due to $CO_2$ adsorption, their $\zeta$-potential is low,[3] and diffusio-phoretic (DP) flows are of negligible magnitude except very close to the IEX surface, where the pH gradients are strongest. We de-carbonized the tracers and charged them up by prolonged batch deionization removing $CO_2$ through ion exchange of its dissociation products (carbonic acid). Further, a reduced substrate $\zeta$-potential was obtained by storing the substrate slides in diluted deconex® cleaning solutions. A stepwise increase in charge ratio had a significant influence on the minimum approach distance. It drastically altered the shape of the formation zone but left the decay unaffected. This is shown in Figure S4a-d. For charge ratios $\chi \leq 1$, we typically observe the single HDW-lines to start directly at the IEX (Figure S4a). As the charge ratio increases, we observe the evolution of sharply bordered, more or less drop-shaped exclusion zones of increasing size (see also Movie S6).

Bright field images and tracer flow fields corresponding to the dark field images in Figures S4a, b, and d are presented in Figure S5. In Figure S5a, we studied the low-charge tracers on a native substrate. Here, the tracer motion directly reflects the underlying DO-flow field. In Figures S5b and c, we used larger charge ratios, but left all other boundary conditions unchanged (tracer concentration, exchange rate, IEX velocity). Therefore, the underlying solvent flow pattern must remain unaltered. In Figures S5b and c, however, we now see a significant alteration of the tracer flow field.

With increasing charge ratio, DP flows gain in strength and importance. Hence, tracers approaching the IEX at its front are stalled already at some distance to the IEX, while tracers initially located in the immediate vicinity of the IEX move outward (Figure S5c). A tracer depleted region forms. Comparison to the bright field images shows that tracers accumulate exactly where outward DP motion and inward DO flow balance and the tracer velocity vanishes: $v_T = v_{DPT} + v_{DO} = 0$. As the IEX passes, the accumulation region first shifts outward, then inward again. Tracers in this boundary are carried along as it shifts inward again (Figure S5b). The evolution of the drop-shaped formation zone in Figures S4a-d is thus seen to originate from changes in the location, where outward DP tracer motion and inward DO solvent flow balance.

Our observations can be rationalized considering that both DP and DO flows originate from the same electrophoretic effect (i.e., from the drop in chemical potential across a charged surface as induced by the increase in the pH of the adjacent solvent) but differ in the relevant type of surface and hydrodynamic boundary conditions. The DP flow originates on the tracer surface, i.e., it reacts to the tracer ζ-potential and the *local* pH gradient. Here, the theoretical concepts of bulk diffusio-phoresis in a quiescent solvent can be directly applied after suitable coordinate transformation.[1,2] By contrast, the DO flow in any volume element above the substrate originates from both the local contribution (depending on the substrate ζ-potential and the local pH gradient) and the *global* hydrodynamic flow pattern. The latter depends on the difference in pH between IEX surface and the background solvent but is also subject to modifications by the hydrodynamic boundary conditions of an extended substrate (and of a closed cell of finite volume). Such a situation is not analytically accessible but may be addressed in numerical calculations.[4] Note that therefore, DP tracer motion follows the local gradient direction and DO solvent motion need not be co-linear.[5] This can clearly be seen comparing the solvent flow field of Figure S5a to the tracer flow fields in Figures S5b and c.

The differences in the position dependence of the two flows become relevant at increased tracer ζ-potentials. Irrespective of $ζ_T$, the tracer DP velocity drops quickly with increasing distance to the IEX, respectively the distance from the trail ridge. As the local pH-gradient along and perpendicular to the trail ridge vanishes with increased distance to the IEX, it becomes negligibly small. By contrast, the DO flows – depending on the large-scale pH differences – are still active far past the IEX. They retain considerable strength, focus the line, and close the initial depletion zone (Figure S5b). A single line results. With increasing tracer ζ-potential, the closing point shifts further away from the IEX. However, the splits always close and a single line is obtained past the line focus as long as a single C-IEX45 is used. By contrast, using a pair of IEX approximately doubles the local pH-gradient while leaving the global pH difference between IEX surface and background unchanged. This shifts the minimum approach distance still further outward. Tracers are now accumulated along a contour far off the IEX center, where the inward DO flow is still very slow (Figure 1, main text). Thus, they can avoid inward transport by the DO flow after IEX passage. The DP split becomes stabilized (Figure S4e).

Quite frequently, an additional line is formed starting at the center of the IEX back (Figures S4f-h). Interestingly, it starts with some delay. Presumably assisted by fluctuations, few tracers reach the IEX, beneath which they are trapped by the solvent flows and expelled once the loading capacity of the IEX is reached. In this type of tracer accumulation at the IEX, the feeding rate is way smaller than by direct DO flow causing the delay. Simultaneously, the roller velocity slightly increases by some 5-10 % and the depleted region slightly stretches. Both indicates a phoretic contribution of the trapped tracers to the propulsion, as well known from modular swimmers.[6] Due to a low feeding rate, this line type typically contributes only little to the merged line in the decay regime.

## 1.4. Alternative ink, alternative pens

The modularity of our approach allows facile testing of the effects introduced by alternative components. By varying the size of silica tracers, both line graininess and contrast can be varied. This is demonstrated in Figure S6 for three species of Si tracers with different sizes as indicated. In all three cases, we used the same tracer concentration $c = 0.1$ wt.%. With decreasing tracer size, both graininess and contrast decrease. For the largest tracers, the lines obtain a "pointillistic" appearance. They become smoother at lower tracer size, but at the same time loose contrast, as scattering power scales with $a_T^{-6}$. Next, we exchanged the silica particles for Polystyrene particles of lower mass density (PS2.3, $\rho_{PS} = 1.05$ kg m$^{-3}$). Three representative images are shown in Figure S7.

The use of lighter tracers generally leads to lines blurred by convection. For tracers of low charge (Figures S7a and b), regions of enhanced density form way above the focusing DO flows along the substrate. For highly charged tracers (Figure S7c), the broadening is additionally enhanced by outward DP motion, which leads to an upward motion already at some distance from the IEX surface. Overall, the use of tracers with low mass density leads to very low contrast and thus, is not useful for writing.

We next checked different types of pens for their capacity to write lines. The examples in Figure S8 feature inert rafts of Si832 tracer particles. These were made by slowly drying Si832 suspensions. The rafts were either free to glide through Si832 or were fixed and passed by the tracers. Short trails resulted, which mainly depended on the shape, orientation, and velocity of the rafts. Figure S8a shows a mobile raft simply ploughing through the tracer layer. Here, a region of enhanced tracer density forms at its front as well as a short tracer-depleted region in its wake. Figure S8b shows a triangular raft fixed in a hydrodynamically favorable orientation. As the tracers drift by, a short trail of slightly increased tracer density forms downstream. Lacking pH gradients, pens made of chemically passive rafts do not trigger DO flows. Line shape and contrast are therefore fully determined by the laminar flow field around these differently shaped and oriented objects. However, in all cases, only very short lines resulted, even at larger gliding velocities. Thus, without attraction, no useful lines are written.

We further replaced the cationic IEX by anionic IEX (A-IEX) and compile the main results in Figure S9. This resin type exchanges residual carbonate ions for OH$^-$. Thus, it increases the pH. A-IEX-L features large exchange rates for $CO_3H^-$ available at concentrations of several µmolL$^{-1}$. They create a pronounced pH variation of complex shape featuring a pH maximum at some distance to the IEX surface (Figure S9a).[5] On its outer side, i.e., pointing away from the IEX, the pH-field is contoured by an inverted gradient (blue arrow in Figure S9b). Small hydrogel-based A-IEX20 spheres show somewhat lower exchange rates and lower capacities. These create a simpler pH field with only negative gradients but of much weaker amplitude and are well suited as pens. Negatively charged tracers approaching from the front are pushed away (both in rolling direction and sideways) by the corresponding *outward* DO flow (Figure S9c). A tracer-depleted region past the A-IEX results. For combination with Si832, we found small AIEX-L splinters to be most suitable. This combination leads to a high-contrast inverted line (Figure S9d). Like the positive lines discussed above, also this line of negative optical contrast appears to be very stable in time (Figure S9e).

### 1.5. Alternative drive

In Figure S10, we again used tracers which are gravitationally bound to the substrate and cationic IEX. We here studied modular micro-swimmers, which propel on a horizontal substrate driven by the DP flow along the surfaces of assembled tracers (see also Movie S7).[6] In all three cases, we observe a characteristic shape of the formation zone, featuring a blob-like region of enhanced tracer density in the immediate back of the IEX. From that blob, a much finer line of tracers is expelled, which only slowly broadens by diffusion. Figure S10a shows a line written by tracers with different sizes past a cationic IEX resin splinter obtained from crushing a larger IEX bead.[7] It demonstrates that a spherical IEX geometry is not essential for writing, nor is a uniform tracer size. Figure S10b shows that, due to mutual repulsion between the negatively charged tracers, the raft past the IEX may take a crystalline structure. However, this has little influence on line formation. In Figure S10c, a modular swimmer is formed by exactly the same components as used in the writing experiments of the main text. Again, a nice line is written past a short formation zone of enhanced tracer density.

Figure S10 demonstrates that writing does not depend on the type of steering chosen. Rather, it requires the presence of an effective attraction (here realized by DO flows). However, modular swimmers lack the ability of precise steering, they steer by statistical rearrangements of their load. Therefore, these freely propelling modular swimmers write curves instead of straight lines or prescribed patterns. In the approach presented in the main text, propulsion by gravity was essential for writing straight lines as prerequisite for drawing more complex patterns.

## 2. Additional Data from Theory

### 2.1 Extracting an effective potential

To model the non-reciprocal effective force between the IEX and the tracer particles, we fitted the tracer velocity field behind the IEX by using different functional forms (Figure S11). The fit was done on the logarithmic scale using the nonlinear least-squares method. The goodness of each fit is measured by the normalized mean-squared distance of the fit values $f_i$ to the data values $d_i$ on the logarithmic scale:

$$\chi^2 = \sum_i \frac{[\ln(d_i) - \ln(f_i)]^2}{\ln(d_i)}. \tag{S1}$$

We found that the functional form given in Equation (1) in the main text provides a good fit to the data. Unlike the other shown fits (dashed-dotted and dotted lines) it also captures the plateau for $r \to 0$ and leads to a reasonable extrapolation capturing the expected power-law decay for $r \to \infty$ (Figure S11).

### 2.2. Reaching the stationary state

In the numerical calculations, the IEX is fixed at $x = 0$ in a homogeneous background of tracer particles at $t = 0$ moving at constant speed $v_{IEX}$. In Figure S12, we show the temporal evolution of the line amplitude in terms of the central probability density $p(t, x, 0)$ on the path of the IEX at certain distances behind the IEX. In Figures S12a-c, we fixed the diffusion coefficient $D = 0.5$ μm$^2$s$^{-1}$ and varied the IEX speed $v_{IEX}$. We obtain two types of curves reaching a stationary state within the investigated time span of 2000 s. At large $v_{IEX}$, a stationary state for $p$ is reached within 100-400 s with a trivial dependence on $x$. The line amplitude plateaus well above the homogenous background. This corresponds to the single-line formation regime of Figure 4 in the main text. At small $v_{IEX}$, a stationary state for $p$ is reached later after some 500-700 s. Notably, the line amplitude reaches a plateau below the homogeneous background. This corresponds to the inverted line formation regime of Figure 4 in the main text. Interestingly, in both types of stationary state, the plateau amplitude decreases with increasing $v_{IEX}$. We attribute this to the smaller number of attracted and/or trapped tracer particles (Figure S2). For speeds $v_{IEX}$ in the transition region between inverted and single lines, the time to reach a stationary state significantly increases because the IEX traps a large amount of tracer particles, which can later escape the trapping region if the number of trapped tracers is large enough. In Figures S12d-f, we fixed the IEX speed to $v_{IEX}$ = 8.6 µms$^{-1}$ and varied the diffusion coefficient. The chosen $x$ are located past the line focus. The plateau amplitude decreases for increasing $D$.

To further study the line amplitude relaxation, we fitted the Gompertz equation

$$p(t) = p_0 \exp\left(-\exp\left[-\frac{t-\Delta t}{\tau}\right]\right) + p_{offset} \tag{S2}$$

with stationary central amplitude $p_0$, time delay $\Delta t$, amplitude relaxation time $\tau$, and background amplitude $p_{offset}$.[8] The fit results are shown in Figure S13. The amplitude relaxation time linearly decreases with the distance $x$ to the IEX (Figure S13a). For a typical distance of 500µm past the IEX, corresponding approximately to the dimension of a single letter, the stationary state is reached with an amplitude relaxation time of $\tau \approx$ 30s. Furthermore, the stationary central amplitude shows a power-law dependence on the distance $x$ (Figure S13b) caused by diffusion processes. The time delay $\Delta t$ of the amplitude relaxation increases linearly with $x$ with a slope equal to the IEX speed (Figure S13c). The time to reach the stationary state hardly changes with varying $D$ and only shows a weak logarithmic dependence (Figure S13d). However, it decreases with increasing IEX speed (Figure S13f). The stationary central amplitude decreases with both the diffusion coefficient and the IEX speed (Figures S13e and g).

## 2.3. Line formation zone

In the line formation zone at very small distances, a split line is observed in the model that merges to a single Gaussian-shaped line at intermediate distances (Figures S14a-c). Past the merging distance, the line first narrows further but eventually broadens as the tracer attraction vanishes and diffusion begins to dominate the tracer motion. The location of this line focus, which separates the line formation zone from the decay zone, defines the line formation length. The latter can be accurately determined via the $x$-dependent standard deviation of Gaussians fitted to the line profiles observed at different IEX speeds (Figure S14). The distance at which the line shows a minimum standard deviation is identified as the line formation length (dashed lines in Figure S14d). The line formation length increases with increasing $v_{IEX}$. For $v_{IEX}$ = 8.6 µm s$^{-1}$, we find $l_f$ = 90 µm in good agreement with the experimental observations.

## 2.4. Discrimination of line types

In the stationary state at $t$ = 2000 s, we fitted single Gaussians and the sum of two equal Gaussians to the line profiles determined at $x$ = -950 µm, i.e., way past the line focus. Figure

S15 shows three examples for the three typical line shapes: inverted line, single line, and split line (see also Figure 4, main text).

## 2.5. Trail formation at early times

Since our minimal model is solely based on the long-range tracer transport towards the IEX and restricted to two dimensions, it is of course not capable to reproduce the detailed tracer motion close to the IEX within the formation zone. In Figure S16, we show the trail formation in the BD simulation with point-like tracer particles at early times. Due to the constriction of the tracer motion to two spatial dimensions, a small layer of tracer particles accumulates at the front and sides of the IEX ($2a \approx 23$ μm, c.f. inset of Figure 3a, main text). Interestingly however, the trail formation behind the IEX is qualitatively similar to the experimental observations. Accumulation at the front is not seen in the experiments because tracer particles can either be swept further underneath the IEX sphere or escape this region by moving along the third spatial dimension due to strong upward solvent flows in the vicinity of the IEX.

## 2.6. Density dependence of the line shape

Thus far, we neglected interactions (volume exclusion) between the tracer particles. Considering repulsive tracer-tracer interaction $u(r)$ modelled by the Weeks-Chandler-Anderson potential[9] with strength $\varepsilon = 10 k_B T_{bath}$ and effective particle diameter $\sigma = 2a_T$ grants interesting insights into the limits and capabilities of our minimal model:

We studied the line shape for interacting tracer particles at different tracer area fractions $\varphi \in \{0.030, 0.076, 0.152, 0.304\}$ which correspond to tracer concentrations of 0.02 wt.%, 0.05 wt.%, 0.10 wt.%, and 0.20 wt.%, respectively. Figure S17 shows simulation snapshots after $t = 110$ s for the different tracer concentrations as well as the density profiles at $x = 500$μm. For higher concentrations, the line width increases. Remarkably, and in contrast to the experimental results (Figure 2d, main text), the line shows a Gaussian shape only at small tracer concentrations. At high concentrations, the shape features a broader, non-Gaussian peak caused by the restriction of the tracer motion to the 2D plane. This contrasts with the model calculations using point-like tracers, which can accumulate to arbitrarily large densities, and to the experiments, where particles may escape to higher elevations. We anticipate that flattening should become observable in experiments using tracers of large density mismatch, which stay settled and cannot escape to the third dimension.

The effects of tracer-tracer interactions have interesting implications on the difference between our experimental and theoretical results: First, the use of point-like tracers in the numerical calculations strengthens the trapping effect because more tracers can be trapped in the vicinity of the IEX as compared to the experiments, in which the tracers have a finite size. Trapping is readily visualized in the present experiments particularly at low IEX velocities (example IV in Figure 4c, main text). However, there, the trapped particles are either sheared off the assembly at its sides, creating a short, diffusively dispersing double line feature, or they are expelled at the back of the assembly, thus filling any underlying depletion zone. As a result, no well-defined density variation (line) emerges, neither under- nor over-density. At larger velocities of $v_{IEX} \approx 8\text{-}10$ μm s$^{-1}$, and in particular, for native substrates of large ζ-potential, the DO flow may be sufficiently strong to create a noticeable underdensity (Figure S4a). However, only a small number of attracted tracers becomes trapped at large velocities. Rather, they are immediately expelled to the back. Thus, we observe a shallow depletion zone to both sides of a strong HDW-line, which extends way beyond the HDW-line focus. The shallow underdensity serves to enhance the optical contrast.

## 3. Accuracy of writing patterns

In the simulations, we prescribe IEX trajectories within the simulation box. Written lines follow these quite accurately, as illustrated in Figure S18. Two small, unavoidable deviations are, however, obvious: First, tracers assembled in the line past the IEX are drawn towards the IEX at its actual position. This originates from the long-ranged nature of the attraction. It leads to an inward offset in the written spiral as compared to the prescribed trajectory (Figure S18a), a right shift in the line crossing event shown in Figure S18b, and a "short-cutting" of turns with small radius of curvature (Figure S18c). In principle, all these effects can be compensated for by appropriate anticipation and fine tuning of the prescribed trajectory (see the writing of the T in TUDa in Movie S5 for instance). The second effect is a local increase in line strength and density upon line crossing events (Figures S18b and c). This local effect occurs, as the IEX acts on the already enhanced density in the immediate environment. Depending on the IEX speed, the corresponding short over-dense region (blob) is shifted in the direction of the effective attraction. The blob occurs past the crossing point. This effect cannot be cured but gets less pronounced for larger velocities.

The line shifts due to attraction are even more pronounced in experiments and increase with increasing substrate charge leading to more pronounced DO flows towards the IEX (c.f. the solvent flow trajectories shown in Figure S5). Blob formation is also present but can be minimized for writing with high IEX velocities $v_{IEX} \geq 10 \mu ms^{-1}$. Yet another effect becomes prominent in experiments attempting to draw complex figures. This is highlighted by yellow encircling in Figure S19. If the pen moves on an already drawn line, its velocity is altered. If it moves against the former propagation direction, it is slowed. This shortens the written line as compared to the programmed trajectory. Moving in the opposite direction, its velocity is increased, and the written line becomes more extended than programmed. Lines are also lengthened if they approach an already written line from an angle. Likewise, lines written such as to pass in the immediate vicinity of an already finished line part get shifted towards the latter. Together, this tends to wreck figures with many line crossings. However, such effects can be accommodated by variation of the programmed stage tilting sequence. Going from Figures S19a-d, we successively adapted the programmed trajectory to make the drawn figure approach the desired one (inset in Figure S19d). While somewhat tedious, this step-by-step procedure is nevertheless successful. In future, it will be improved by replacing the manual programming by some learning package based on image processing feedback.

## 4. Blotted lines

Blob formation upon line-crossing has been discussed in the main text. Occasionally, we observe an additional type of blob formation. It is caused by a transient sticking of the IEX to the substrate. This process of blob formation is followed in a series of images in Figure S20. The blob is mainly formed by the trail catching up with the now-stuck IEX and further tracer accretion at the fixed IEX. Note that during sticking, the tracers slowly overtake the IEX. After detachment, the IEX velocity initially is slower than the final velocity. This is attributed to IEX friction with the locally enhanced tracer density but may further be related to altered DO flows. By contrast to blobbing upon line crossing, this type is not systematic. It therefore may be an issue already in straight line writing. Blobbing by transient sticking is most prominent for low-charge substrates and elevated tracer concentrations. It is also more frequent at lower velocities. Blob formation events are uncorrelated, and the resulting line appears to be blotted (Figure S20d). A regular blobbing pattern might be introduced for decorative purposes by transiently trapping the IEX by an optical tweezer for example.

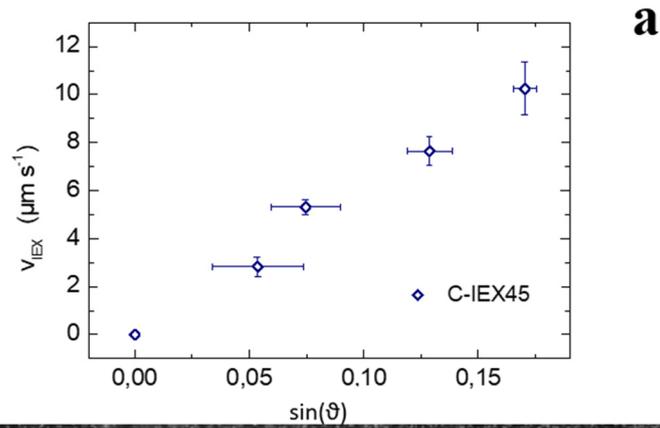
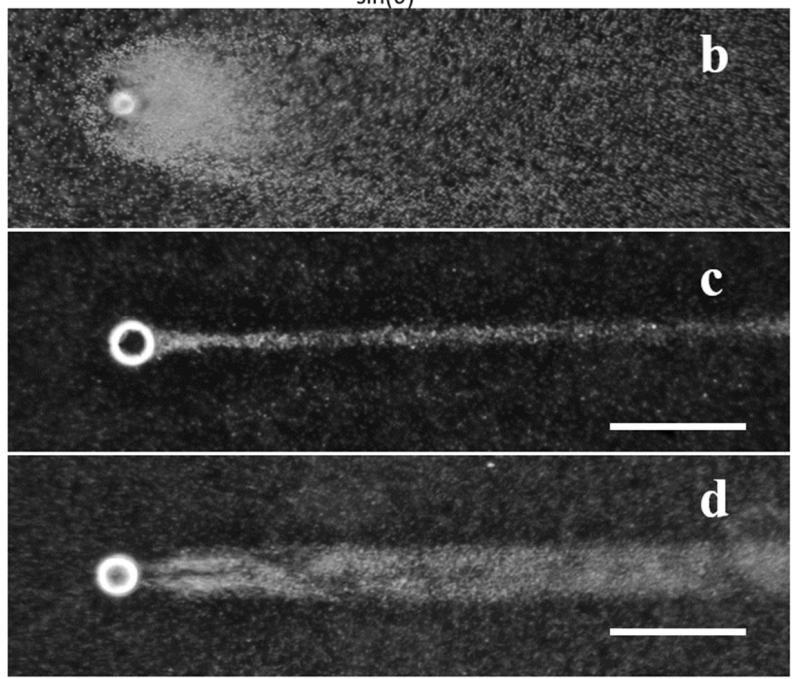

**Figure S1.** Tuning the inclination. a) Speed of C-IEX45 rolling through tracer free water as a function of substrate inclination. b-d) Dark filed images taken at different velocities of C-IEX45. Scale bars: 200 µm. b) C-IEX45 rolling through 0.1 wt.% Si2.1 at very low $v_{IEX} \approx$ 0.28 µm s$^{-1}$. Note the extended asymmetric accretion zone. No useful line is written. c) C-IEX45 rolling through 0.1 wt.% Si832 at $v_{IEX}$ = 7.2 µm s$^{-1}$. A straight, narrow line is obtained. d) C-IEX45 rolling through 0.16 wt.% Si832 at $v_{IEX}$ = 14 µm s$^{-1}$. Note the initial line split, which quickly washes out due to diffusion.

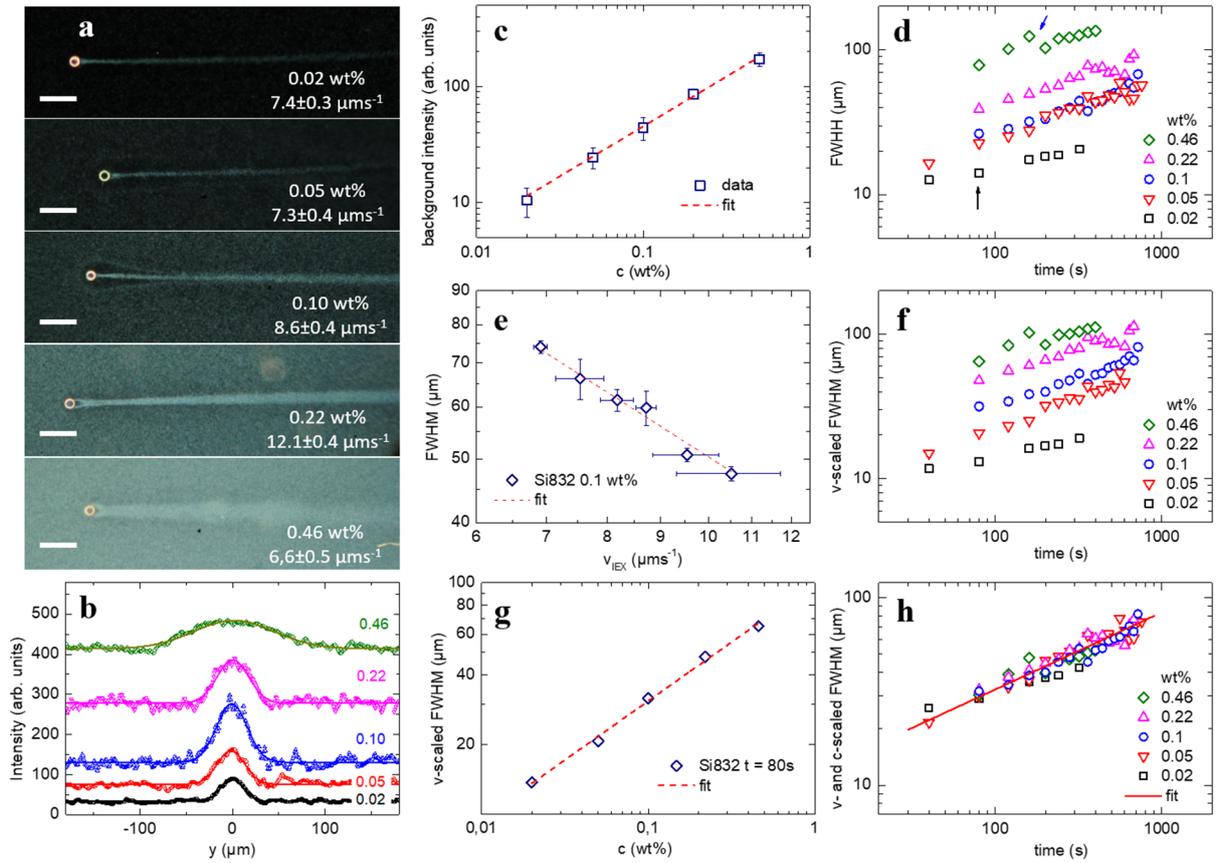

**Figure S2.** Line width measurements. a) Dark field snapshots of C-IEX45 rolling with speeds in the range of 6-12 µm through Si832 of different concentrations as indicated. Scale bars: 200 µm. Concentrations and speeds are indicated in each image. b) Line profiles recorded 80 s after IEX passage. The solid lines are fits of a Gaussian with an offset corresponding to the background scattering intensity. c) Double-logarithmic plot of the concentration dependent scattered intensity in dark field images of tracers equilibrated on a horizontal substrate. The fit of a linear function to the data returned a power law exponent of $a = 0.85 \pm 0.04$. d) Double-logarithmic plot of the time dependent FWHM for the five experiments shown in panel (a). e) Double-logarithmic plot of the dependence of line width on IEX speed for C-IEX45 rolling through Si832 at 0.18 wt.%. The fit of a linear function to the data (dashed line) returns a power law exponent of $-1.02 \pm 0.09$. f) Double-logarithmic plot of the time-dependent FWHM scaled to $v_{IEX} = 8$ µm s$^{-1}$. g) Double-logarithmic plot of the $v$-scaled FWHM in dependence on Si832 concentration. The fit of a linear function to the data (dashed line) returns a power law exponent of $a = 0.51 \pm 0.02$. h) Double-logarithmic plot of the time-dependent FWHM scaled to $v_{IEX} = 8$ µm s$^{-1}$ and $c = 0.1$ wt.%. The fit of a linear function to the complete data (solid line) returns a power law exponent of $\lambda = 0.48 \pm 0.04$.

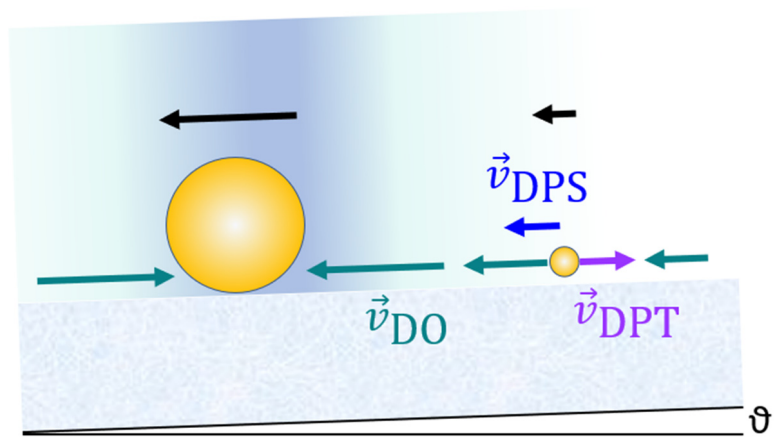

**Figure S3.** Relevant velocities during tracer approach. Due to gravity, both IEX and tracers roll at speeds $v_i$ down the substrate tilted by $\vartheta$ (black arrows). The incompressible solvent flows with local velocities $v_{DO}(x,y)$ under the influence of the pH field (indicated by the background shading). The DO flow accelerates as it converges at the IEX (green arrows). The pH-gradient further induces a diffusio-phoretic (DP) slip at the charged tracer surfaces. This results in an inward solvent flow (DPS, dark blue arrow) relative to the tracer surface and an outward tracer motion (DPT, violet arrow).

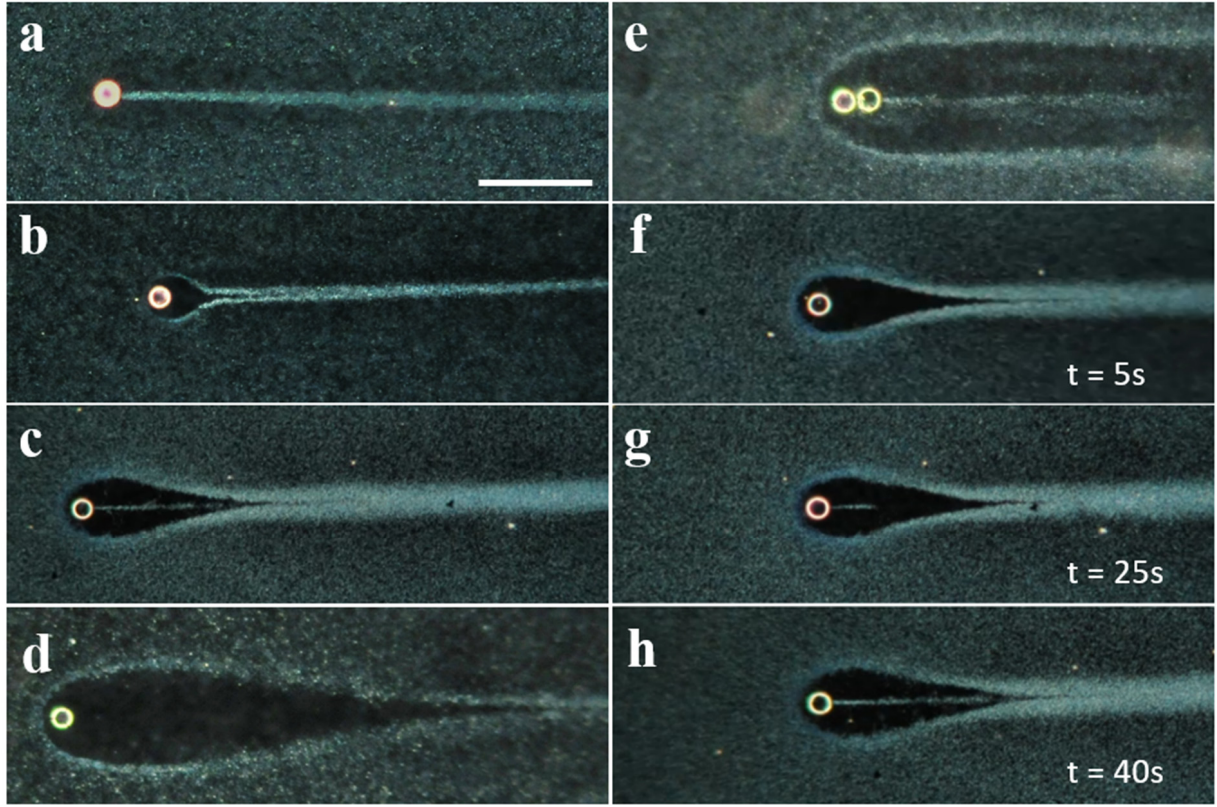

**Figure S4.** Manipulating the shape of the line formation zone. Dark field images of CIEX45 rolling through Si832 at $\vartheta = 3.1$ ($v_{IEX} = 7.7$ μm s$^{-1}$). Scale bars: 200 μm. We here changed the minimum approach distance by varying the charge ratio $\chi$. a) 0.02 wt.% Si832, salt free but non-decarbonized ($\zeta_T = -68$ mV) on a native substrate of $\zeta_S = -105$ mV. Note the slight depletion of Si832 from the immediate IEX surroundings. b) Same as in a, but after de-carbonizing Si832 for 1h ($\zeta_T = -80$ mV) and on a deconexed substrate ($\zeta_S = -70$ mV). c) Same as in panel (a), but after de-carbonizing Si832 for 24h ($\zeta_T = -102$ mV) on a deconexed substrate ($\zeta_S = -70$ mV). d) Single IEX rolling in 0.16 wt.% Si832 deionized for 21d ($\zeta_T = -108$ mV) on a deconexed substrate ($\zeta_S = -70$ mV). e) IEX-pair rolling through 0.16 wt.% Si832 deionized for 21d ($\zeta_T = -108$ mV). f-h) Temporal development of the pattern seen in c. Snapshots were taken at different times after the start of the experiment as indicated in the key.

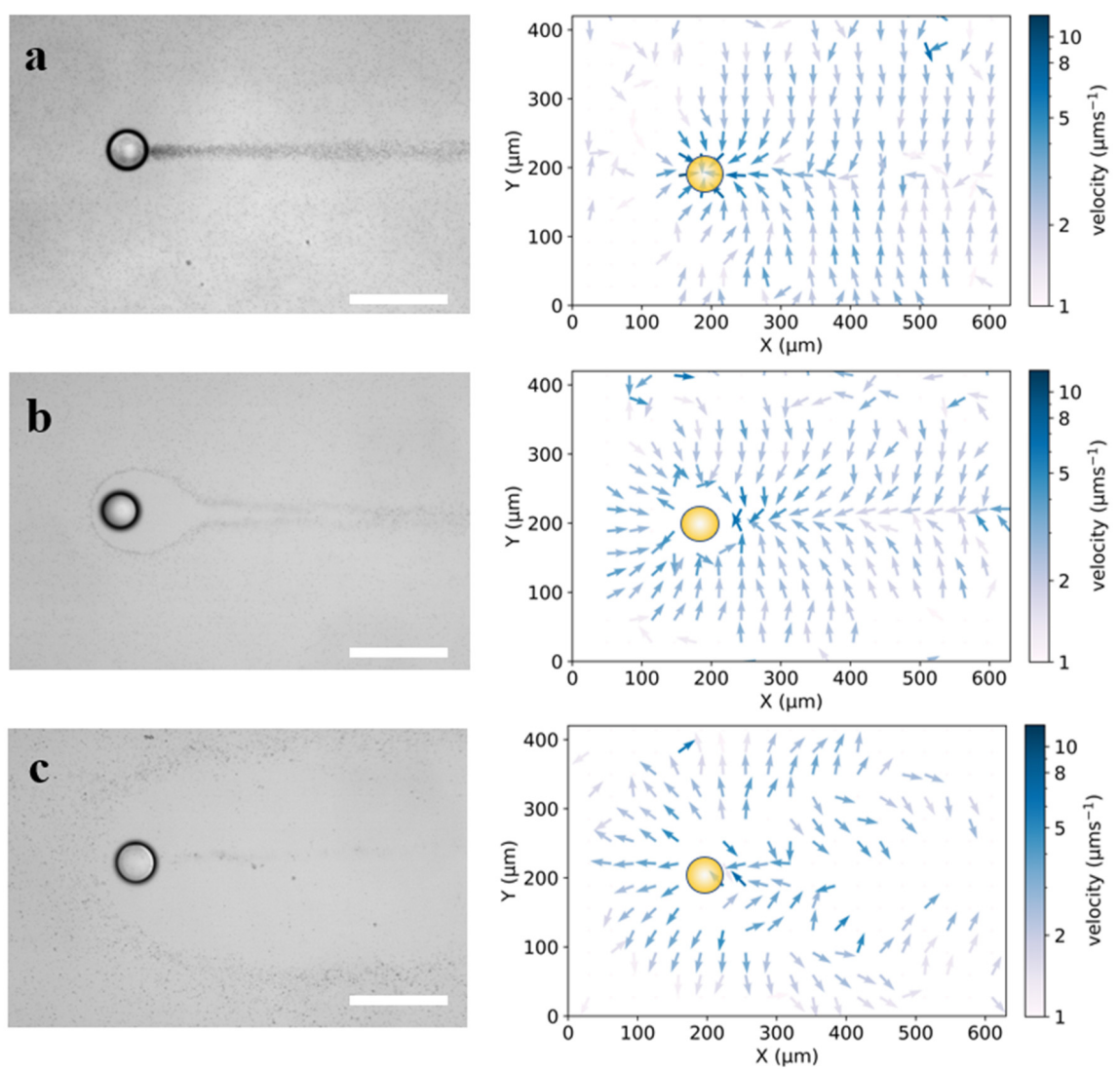

**Figure S5.** Flow analysis. Bright field images (left, scale bar 100 µm) and tracer flow fields (right) for C-IEX45 rolling at 3.1° inclination ($v_{IEX}$ = 7.7 µm s$^{-1}$). The flow direction is given by the orientation of the arrows and the instantaneous lab-frame velocity is encoded by the arrow color as indicated in the key. Images and maps were recorded for C-IEX45 rolling on differently charged glass substrates through 0.10 wt.% Si832 tracers. a) Salt free but not de-carbonized tracers ($\zeta_T$ = -68 mV) on a native substrate of $\zeta_S$ = -105 mV (same conditions as in Figure S4a). Under these conditions, DP flows are negligible and the tracer flow field coincides with the solvent flow field. b) Si832 deionized for 1h ($\zeta_T$ = -80 mV) on a deconexed substrate of $\zeta_S$ = -70 mV (same conditions as in Figure S4b). The tracer flow field shows significant differences when compared to the previous situation. c) Si832 deionized for 21d ($\zeta_S$ = -108 mV) on a deconexed substrate of $\zeta_S$ = -70 mV (same conditions as in Figure S4d). Note the now extended region of outward tracer motion next to the IEX in the velocity maps.

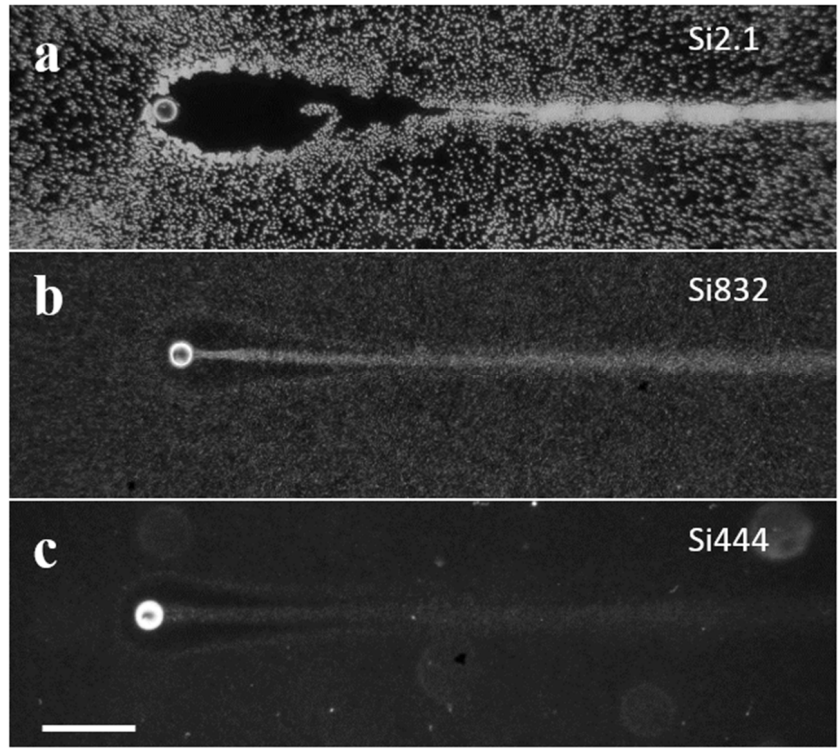

**Figure S6.** Variation of graininess and optical line contrast. Moderately de-carbonized C-IEX45 rolling at $v_{IEX}$ = 8.4 µm s$^{-1}$ on low-charge substrates. Scale bar: 250 µm. a) Si2.1 at $c$ = 0.1 wt.%. b) Si832 at $c$ = 0.1 wt.%. c) Si444 at $c$ = 0.1 wt.%.

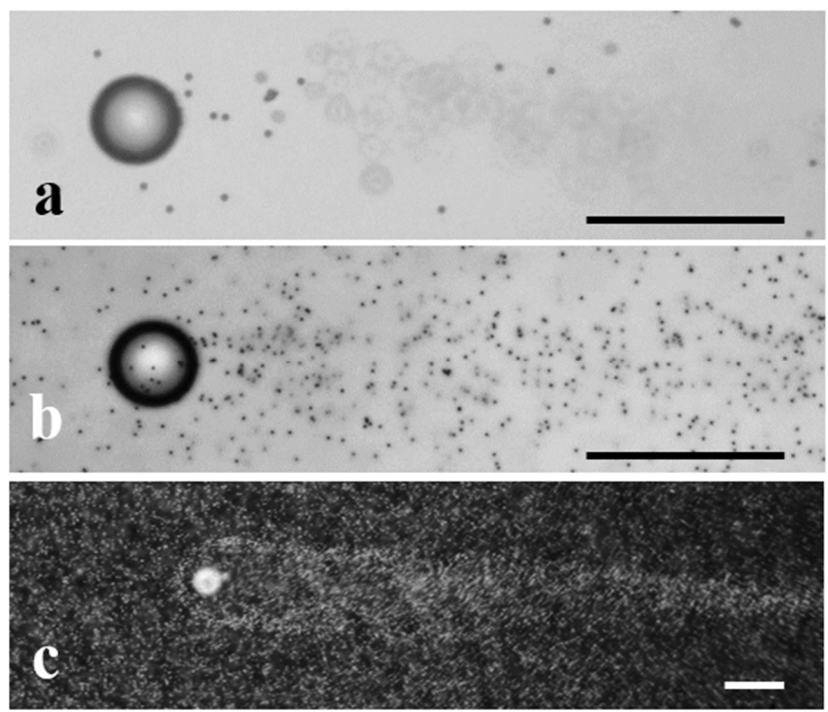

**Figure S7.** Weak-contrast lines written by C-IEX45 in different suspensions of light tracers. Scale bars: 100 μm. a) Bright field image of C-IEX45 rolling on an inclined substrate ($\vartheta = 2.3°$) in a dilute suspension of PS2.3 ($\zeta_T = -35$ mV). The tracers in focus approach the C-IEX along the substrate. The actual trail forms above (blurred tracer images). b) The same situation but now focused to some 150 μm above the substrate. c) Dark field images of C-IEX45 rolling on an inclined substrate ($\vartheta = 2.3°$) in suspension of decarbonized PS2.3 ($\zeta_T = -65$ mV) at 0.10 wt.%.

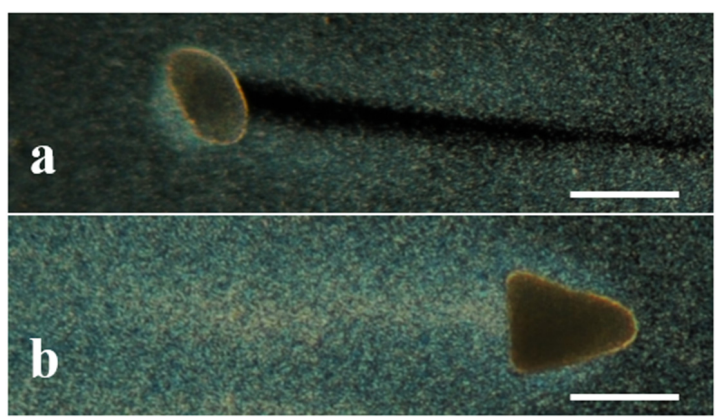

**Figure S8.** Writing with inert objects. Dark field images of weak trails formed by inert objects moving through suspensions of tracer particles. Scale bars: 100 μm. a) Chemically inert raft of dried Si832 gliding down an inclined substrate ($\vartheta = 7.6°$) in a suspension of Si832 at 0.10 wt.%. b) Chemically inert raft of dried Si832 fixed to an inclined substrate ($\vartheta = 7.6°$) in a suspension of Si832 at 0.10 wt.%.

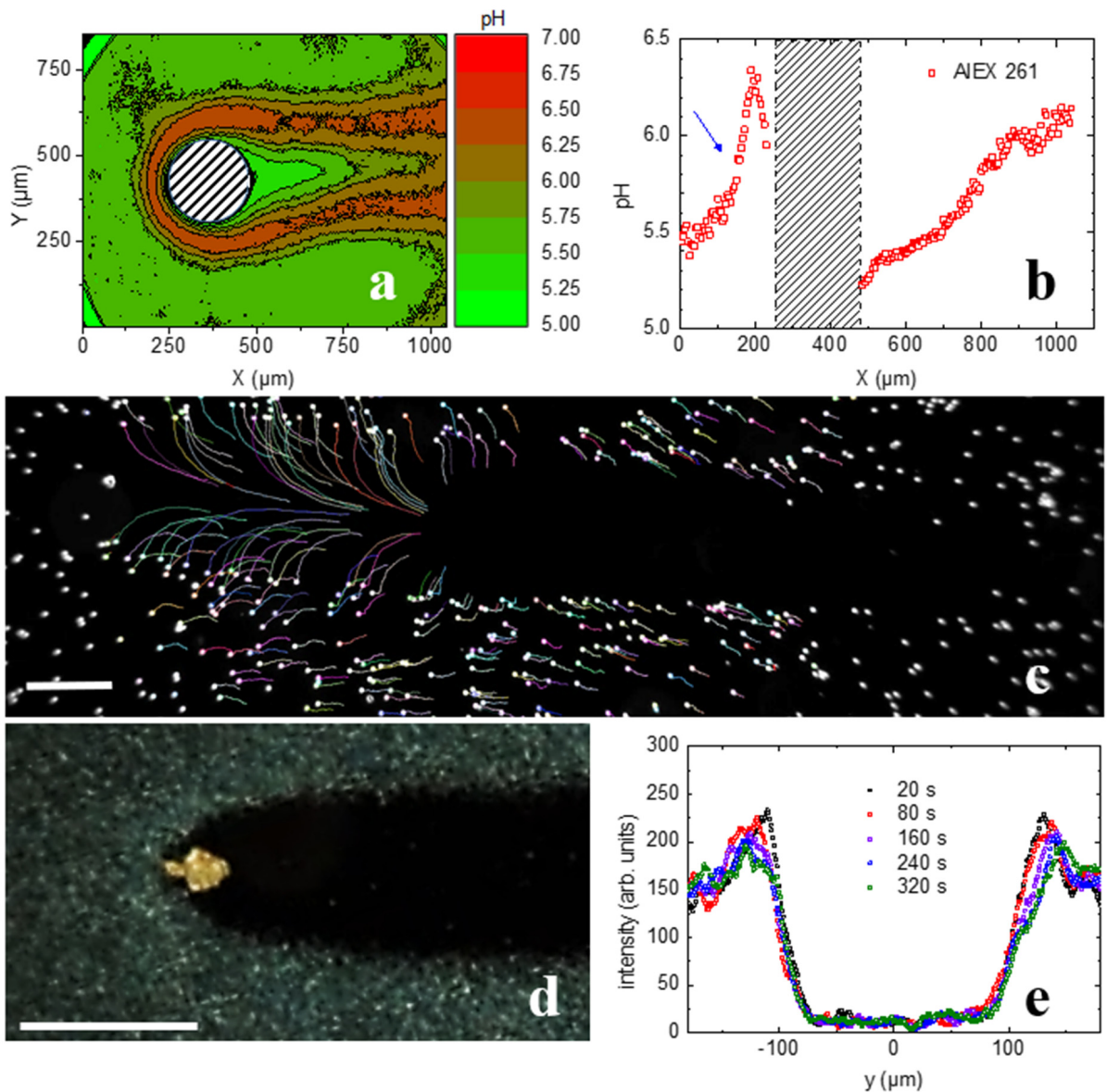

**Figure S9.** Contrast inversion by inversion of the pH gradient direction. a) pH-map of a resin-type anionic A-IEX-L of 261 µm diameter. Due to the high exchange rate, a complex pH pattern evolves. b) pH profile recorded along the rolling direction. The inverted gradient to the front of the AIEX is indicated by the blue arrow. Note however, that the whole pH-field is bordered by such an outward-decreasing pH field in panel (a). c) Tracer trajectories recorded in darkfield for a hydrogel-based A-IEX15. Scale bar: 200 µm. The ensuing pH filed is similar but less pronounced for this smaller resin species. Tracers are swept away from the A-IEX in forward direction and sideways. d) Small resin-type A-IEX-L splinter gliding down an inclined substrate in Si832 at 0.1 wt.% and creating a line of negative optical contrast. Scale bar: 250 µm. e) Line profiles of the inverted line drawn in (d) for different times after A-IEX passage as indicated.

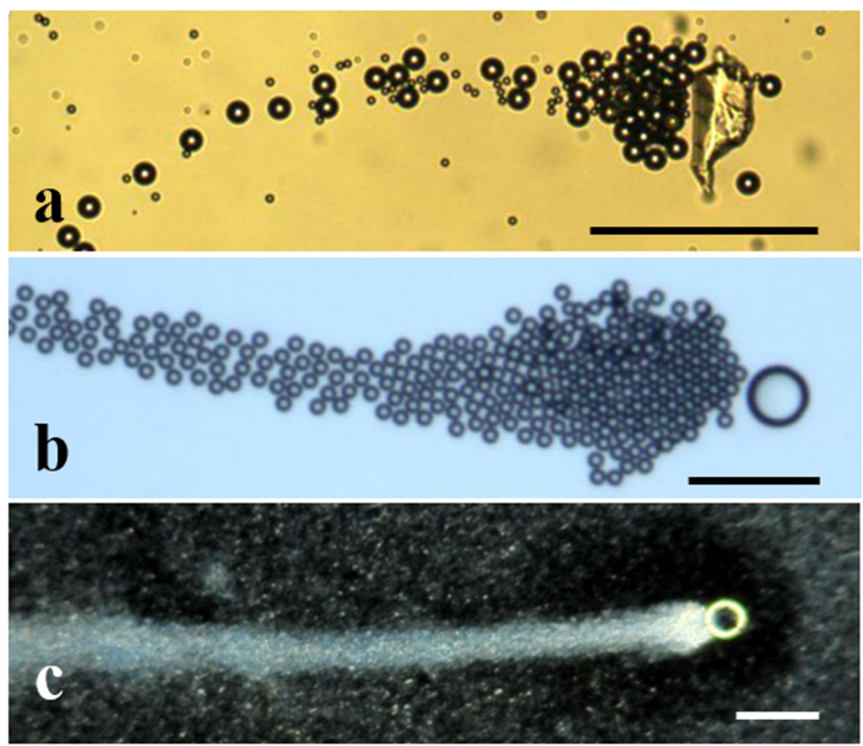

**Figure S10.** Writing with self-propelling pens. Images of lines formed by different cationic IEX types on horizontal native substrates in different tracer suspensions. These objects move as modular phoretic swimmers. Scale bars: 100 µm. a) Swimmer formed by a resin type C-IEX splinter in a dilute binary suspension of PS15, PS5.2 and PS1.7. b) Swimmer formed by microgel-type C-IEX45 in a dilute suspension of PS10. c) Swimmer formed by C-IEX45 in a suspension of Si832 at 0.10 wt.% on a horizontal substrate. Note the diffusely bordered depletion zone close to the C-IEX demonstrating the effect of the DO flows.

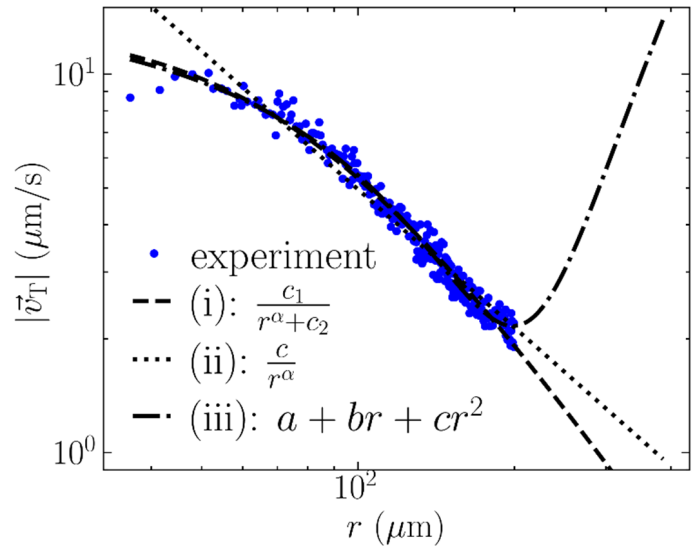

**Figure S11.** Nonlinear least-squares fits of the velocity field behind the IEX as used to extract a functional form of the effective interaction between IEX and tracers. The data is fitted on the logarithmic scale. The three shown fit functions (i)-(iii) have goodness 0.94, 1.55, and 0.81, respectively [see Equation (S1)]. Since fit (iii) results in an unphysical behavior at large distances and the simpler two-parameter fit (ii) has a significantly worse goodness, we finally used fit (i) which captures both the plateau at small *r* and the power-law decay at large *r*.

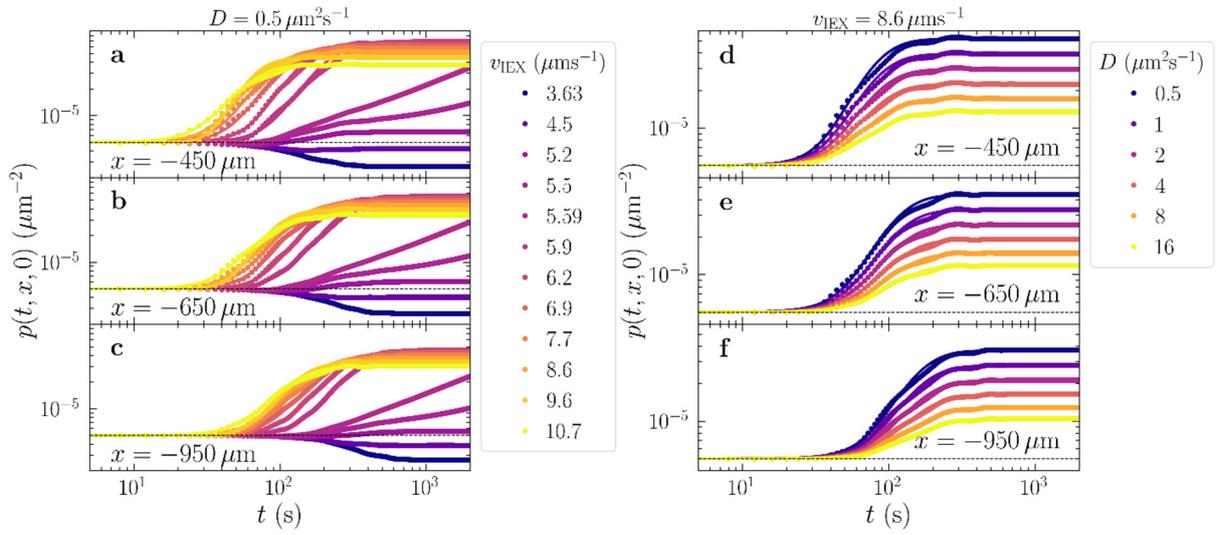

**Figure S12.** Time evolution of the line amplitude. Central probability density $p(t < 2000, x, 0)$ at different distances $x$ from the IEX (values are given in the panels) as obtained from the numerical solution of Equation (6) in the main text. The solid lines are fits of Equation (S2) to the data. a-c) Data obtained for fixed diffusion constant $D = 0.5$ µm² s⁻¹ and different IEX speeds $v_{IEX}$ (color-coded according to the key). d-f) Data obtained for fixed IEX speed $v_{IEX} = 8.6$ µm s⁻¹ and different diffusion coefficients $D$ (color-coded according to the key). Other parameters as given in Table 2 in the main text.

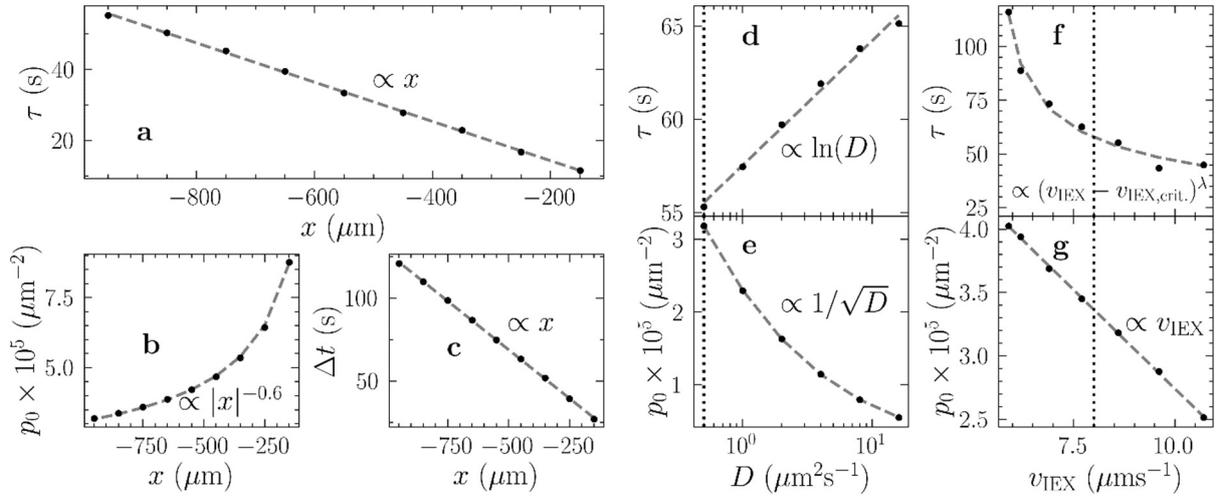

**Figure S13.** Analysis of the time evolution of the line amplitude as obtained from the fits shown in Figure S12. Black dots correspond to the respective fit parameters, grey dashed lines to fits as given in the key. a) Amplitude relaxation time $\tau$, b) stationary central amplitude $p_0$, and c) time delay $\Delta t$ over distance $x$ to the IEX for $D = 0.5$ µm² s⁻¹ and $v_{IEX} = 8.6$ µm s⁻¹. d) Amplitude relaxation time $\tau$ and e) stationary central amplitude $p_0$ as function of the diffusion coefficient for fixed $v_{IEX} = 8.6$ µm s⁻¹ at $x = -950$ µm. f) Amplitude relaxation time $\tau$ and g) stationary central amplitude as function of the IEX speed for fixed $D = 0.5$ µm² s⁻¹ at $x = -950$ µm. The critical IEX velocity $v_{IEX,crit.}$ denotes the transition point from inverted to single lines. Vertical dotted lines denote typical values as used in the experiment.

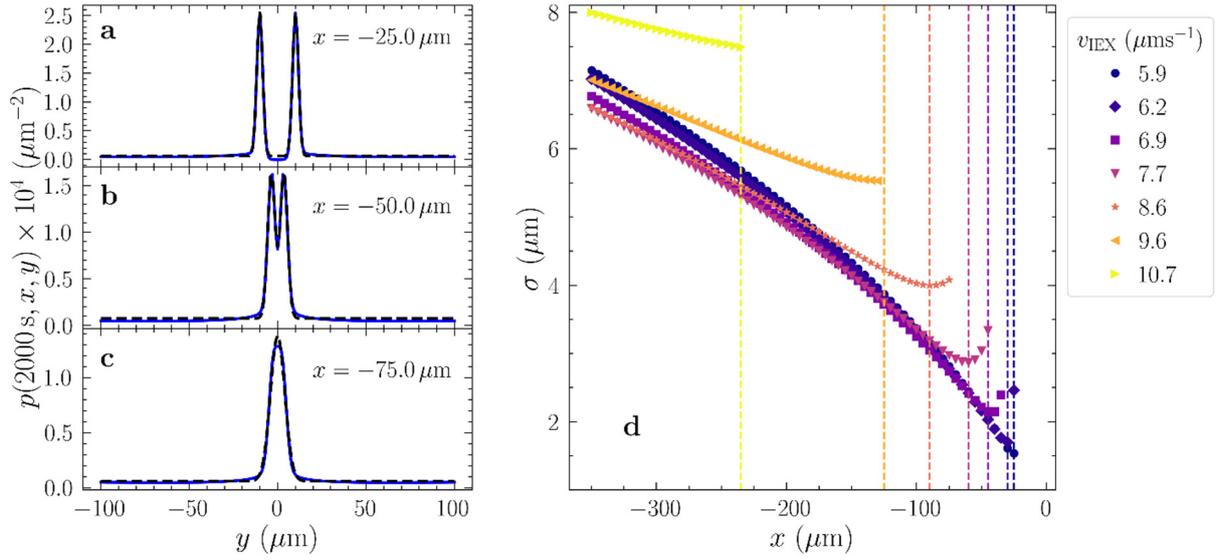

**Figure S14.** Determination of the line formation length. a-c) Line profiles for $v_{IEX}$ = 8.6 µm s$^{-1}$ and $D$ = 0.5 µm$^2$ s$^{-1}$ and d) line width for $D$ = 0.5 µm$^2$ s$^{-1}$ and different $v_{IEX}$ obtained from the numerical solution of Equation (6) in the main text as a function of the distance past the IEX. The dashed black lines in panels (a)-(c) are fits to one Gaussian and a sum of two Gaussians, respectively. In panel (d), we show the standard deviation of the Gaussians fitted to the line profiles for different speeds $v_{IEX}$ of the IEX (values are given in the key). A minimum coincides with the location of the line focus, whose position we denote as the line formation length (dashed vertical lines). For $x$ closer to 0, the line is not a single Gaussian anymore but rather a split line. Therefore, the curves stop at certain $x$.

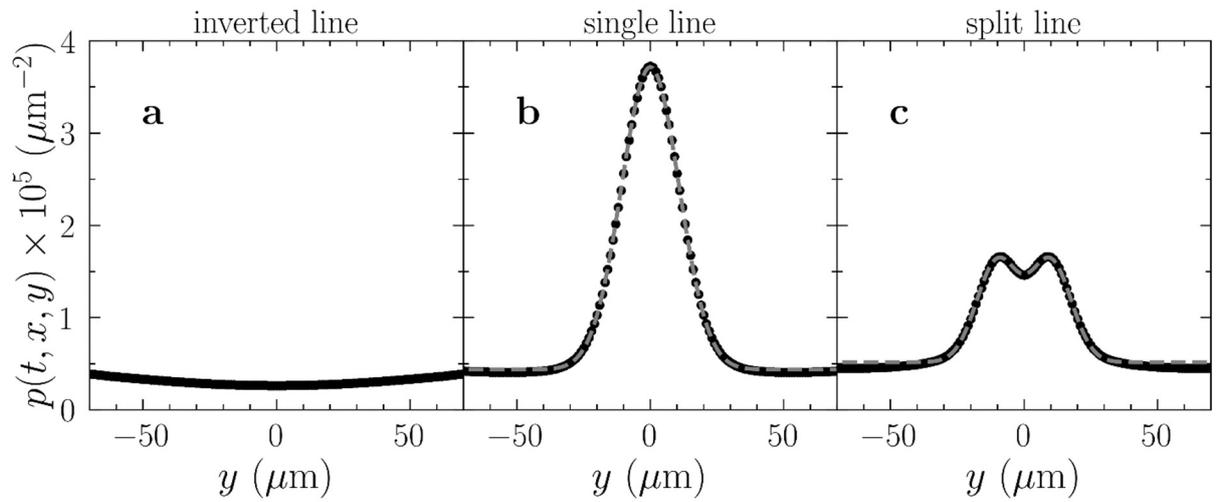

**Figure S15.** Analysis of line shapes. Exemplary line profiles obtained in the continuum model (Equation (6) in the main text) at $t = 2000$ s and $x = -950$ μm with $D = 0.5$ μm$^2$ s$^{-1}$. The dashed lines denote the corresponding fits of single or double Gaussians. a) $v_{IEX} = 3.63$ μm s$^{-1}$, b) $v_{IEX} = 8.6$ μm s$^{-1}$, and c) $v_{IEX} = 16.4$ μm s$^{-1}$. Note the pronounced differences in line amplitudes. All other parameters are given in Table 2 in the main text.

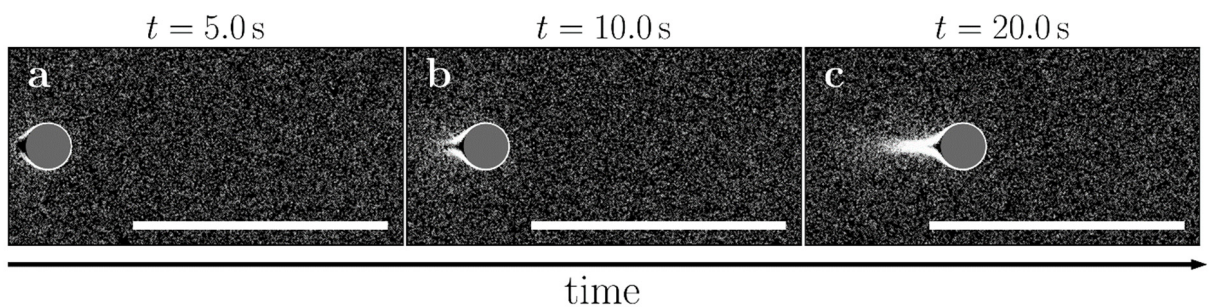

**Figure S16.** Short-time trail formation. Onset of writing for non-interacting tracer particles of radius $a_T$ = 0.416 μm and diffusion coefficient $D$ = 0.5 μm² s⁻¹ at three different (early) times (a-c) ordered from left to right (values are given in the panels). The IEX moves at a speed $v_{IEX}$ = 8.0 μm s⁻¹ and the area fraction of tracer particles is $\varphi$ = 0.152. All other parameters as given in Table 2 of the main text. Scale bar: 250 μm.

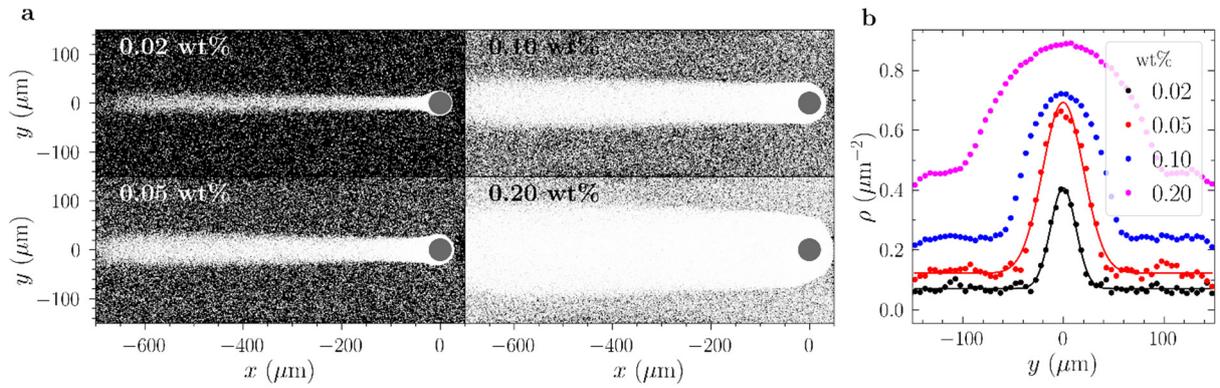

**Figure S17.** Density dependence of line profiles in Brownian dynamics simulations with repulsively interacting tracer particles of radius $a_T$ = 0.416 μm and diffusion coefficient $D$ = 0.5 μm$^2$ s$^{-1}$. The IEX moves at a speed $v_{IEX}$ = 8.0 μm s$^{-1}$. a) Simulation snapshots after $t$ = 110 s taken at different tracer concentrations (i.e., different area fractions) as indicated by the key. b) Corresponding line profiles obtained from averages over regions of 50 μm extension at a distance of $x$ = -500 μm past the IEX. Solid lines are Gaussian fits, describing the data well at low tracer concentrations (i.e., area fractions). Other parameters are given in Table 2 in the main text.

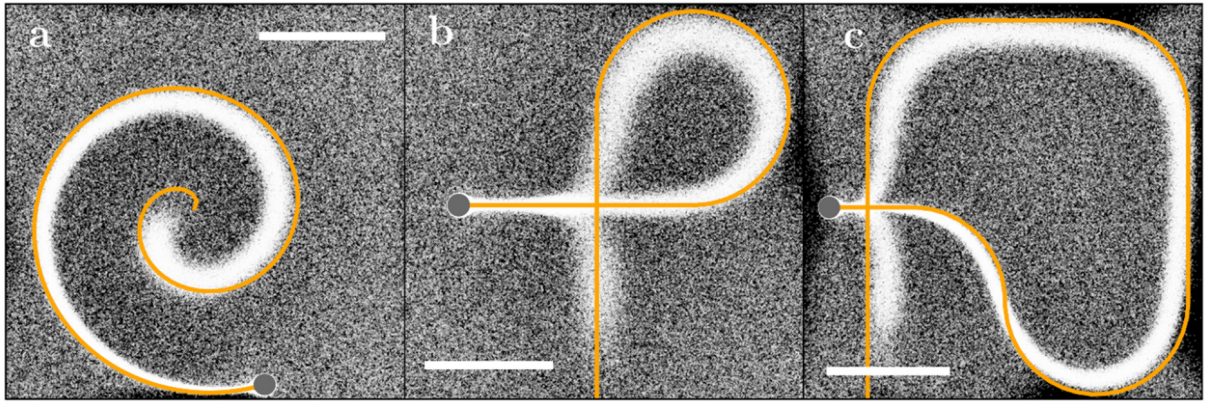

**Figure S18**. Exemplary patterns drawn in the BD simulation. Scale bar: 250 µm. a) Spiral. b) Clockwise loop with line crossing. c) The same as in panel (b) but crossing starting after an additional left turn. Note the blob just past the crossing and the slight deformation of the already drawn line. Simulation parameters are given in Table 2 in the main text.

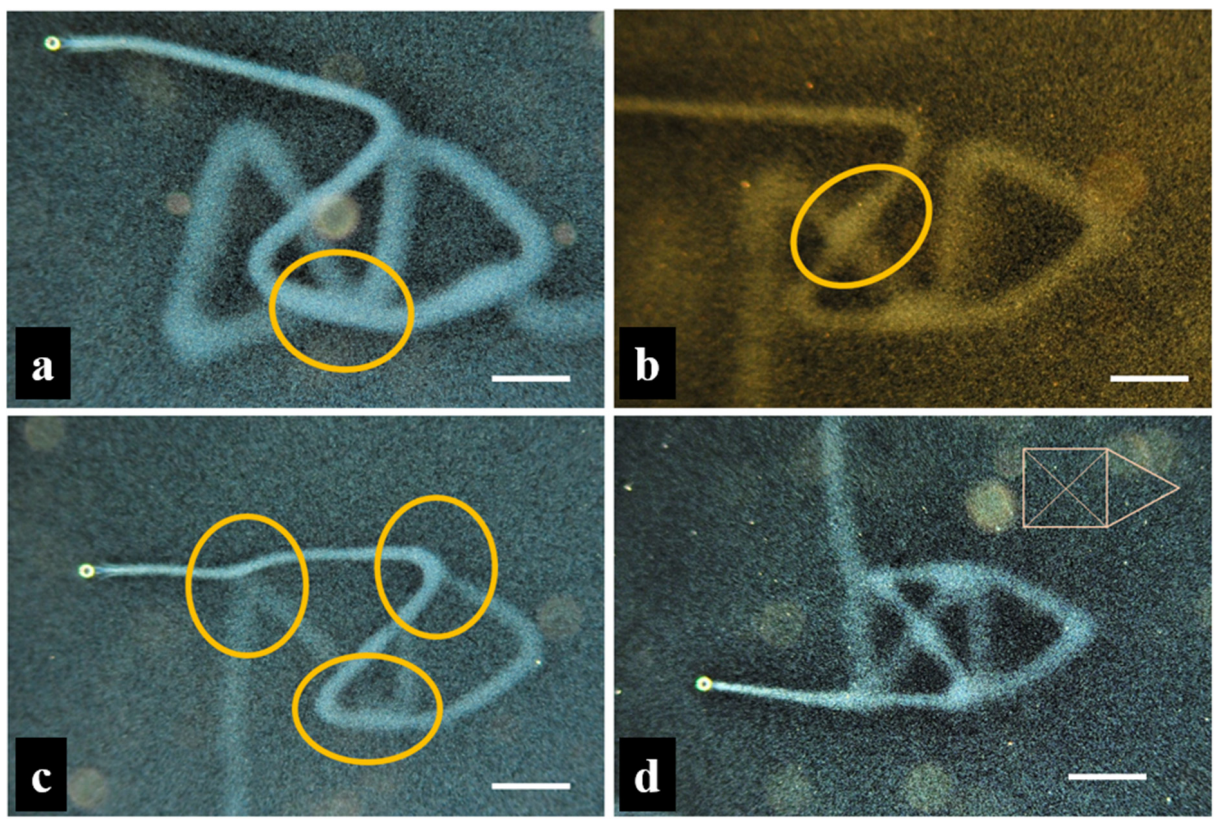

**Figure S19.** Approaching a perfect drawing in experiment. By trial and error, we successively improved the adjustment of the stage programming. Scale bar: 250 μm. a-c) Results obtained at intermediate stages. The encircled regions highlight deviations from the ideal pattern shape (line length variations, lateral line shifts, line bending, and blobbing, see Supporting Text). d) Final result.

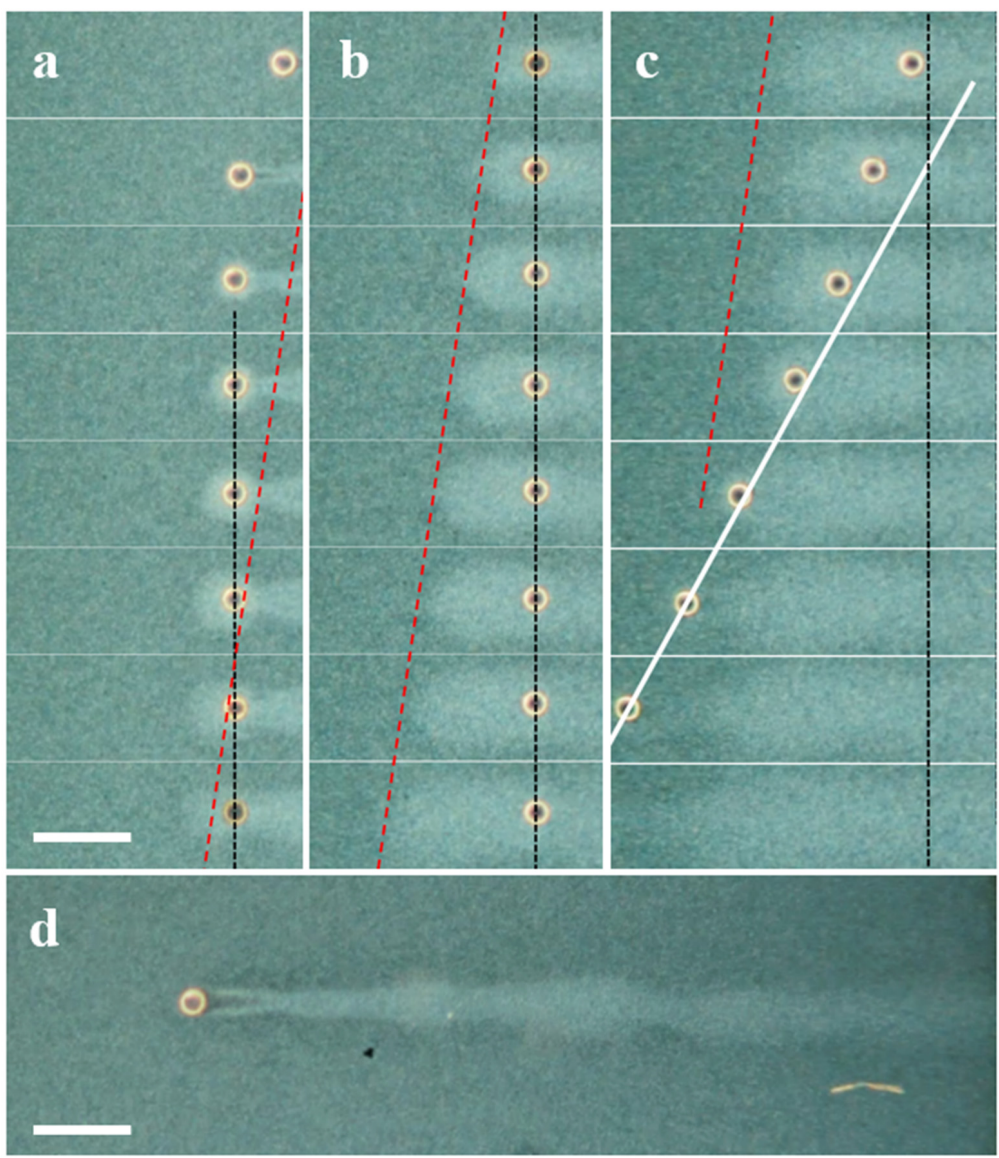

**Figure S20.** Formation of a blotted line. Time series of dark field images of blob formation due to a transient sticking of the IEX to the substrate. C-IEX45 rolling through Si832 at $c =$ 0.2 wt.%. Scale bars: 250 μm. a-c) Development of a blob. Time increases from top to bottom and from panel (a) to (c). Images are separated by 800 ms. The dashed black line marks the position at which the IEX stuck to the substrate. The red line is a guide to the eye, marking the location of the blob tip, which continues to move leftward at roughly constant velocity, until the IEX is detached again. The solid white line denoted the constant velocity of the IEX after escaping the blob. Note the acceleration stage immediately after detachment. d) Frequent sticking results in blotted lines.

**Movie S1.**

CIEX45 in 0.1 wt.% native Si832 on a native substrate. Focus on line formation.

**Movie S2.**

CIEX45 in 0.1 wt.% native Si832 on a native substrate. Focus on line stability/decay.

**Movie S3.**

Brownian dynamics simulation of writing a straight line. Parameters as given in Tab. 2 in the main text.

**Movie S4.**

Numerical solution of the continuum model [Equation (6) in the main text]. Parameters as given in Tab. 2 in the main text but with $v_{IEX} = 8.6\,\mu\text{m s}^{-1}$.

**Movie S5.**

Writing "TUDa" in the Brownian dynamics simulation at $v_{IEX} = 12\,\mu\text{m s}^{-1}$. Other parameters as given in Tab. 2 in the main text.

**Movie S6.**

CIEX45 in 0.1 wt.% 2h deionized Si832 on a deconexed substrate. Focus on DP effects and changed flow field within the formation zone.

**Movie S7.**

Demonstration of modularity – writing with a self-propelling pen. CIEX45 in 0.1 wt.% native Si832 on a horizontal native substrate.

**Movie S8.**

Demonstration of erasing written letters by globally heating the system and recycling the ink in a Brownian dynamics simulation at $v_{IEX} = 12\,\mu\text{m s}^{-1}$. Other parameters as given in Tab. 2 in the main text. Scale bar: 250 μm.

**Movie S9.**

Demonstration of erasing a written line and rewriting at a shifted position by collecting the initial line with a second IEX following the "corrected" trajectory. CIEX45 in 0.1 wt.% Si832.